\documentclass[11pt]{article}

\usepackage{lscape}
\usepackage{graphicx}
\usepackage{rotating}
\usepackage{authblk}
\usepackage{amsmath}
\usepackage{xcolor}
\usepackage{soul}
\usepackage{multirow}
\usepackage{amssymb}
\usepackage{enumitem}
\usepackage{tikz}

\usepackage{blindtext}

\title{Bivariate network meta-analysis for surrogate endpoint evaluation}

\author[1]{Sylwia Bujkiewicz*}
\author[2]{Dan Jackson}
\author[3]{John R Thompson}
\author[4]{Rebecca Turner}
\author[1]{\\Keith R Abrams}
\author[4]{, and Ian R White}

\affil[1]{Biostatistics Research Group, Department of Health Sciences,
            University of Leicester, U.K.}
\affil[*]{email: sylwia.bujkiewicz@le.ac.uk}

\affil[2]{Statistical Innovation Group, Astrazeneca, U.K.}

\affil[3]{Genetic Epidemiology Group, Department of Health Sciences, University of Leicester, U.K.}

\affil[4]{MRC Clinical Trials Unit, University College London, U.K.}

\date{7 March 2018}

\begin{document}

\maketitle

\abstract{Surrogate endpoints are very important in regulatory decision-making in healthcare, in particular if they can be measured early compared to the long-term final clinical outcome and act as good predictors of clinical benefit. Bivariate meta-analysis methods can be used to evaluate surrogate endpoints and to predict the treatment effect on the final outcome from the treatment effect measured on a surrogate endpoint. However, candidate surrogate endpoints are often imperfect, and the level of association between the treatment effects on the surrogate and final outcomes may vary between treatments. This imposes a limitation on the pairwise methods which do not differentiate between the treatments. We develop bivariate network meta-analysis (bvNMA) methods which combine data on treatment effects on the surrogate and final outcomes, from trials investigating heterogeneous treatment contrasts. The bvNMA methods estimate the effects on both outcomes for all treatment contrasts individually in a single analysis. At the same time, they allow us to model the surrogacy patterns across multiple trials (different populations) within a treatment contrast and across treatment contrasts, thus enabling predictions of the treatment effect on the final outcome for a new study in a new population or investigating a new treatment. Modelling assumptions about the between-studies heterogeneity and the network consistency, and their impact on predictions, are investigated using simulated data and an illustrative example in advanced colorectal cancer. When the strength of the surrogate relationships varies across treatment contrasts, bvNMA has the advantage of identifying treatments for which surrogacy holds, thus leading to better predictions.}

\section{Introduction}

Surrogate endpoints are very important in the drug development process, at both the trial design and the evaluation stage.
They are particularly useful when they can provide early measurement of the treatment effect, in settings where a long follow up time is required before measurement of the final clinical outcome  \cite{burzykowski2006}.
This is often the case in cancer where overall survival (OS) is of primary interest whilst other outcomes such as progression-free survival (PFS) potentially can be used to measure the effect of a treatment earlier.
Alternatively, PFS may be of primary interest and tumour response (TR) is then investigated as a short term surrogate endpoint to PFS.
Before they can be used in evaluation of new health technologies, candidate surrogate endpoints have to be assessed for their predictive value of the treatment effect on the final clinical outcome.
Surrogate outcomes are validated by estimating the pattern of association between the treatment effects on surrogate and final endpoints across trials using meta-analytic techniques \cite{daniels1997, buyse2000, burzykowski2001, renfro2012, bujkiewicz2015, bujkiewicz2016}.

Multivariate meta-analysis methods are used to obtain average treatment effects on multiple endpoints while taking account of the correlations between them \cite{vanhouwelingen2002, riley2007, wei2013, bujkiewicz2013} and, as such, are suitable tools for modelling surrogate endpoints \cite{bujkiewicz2015, bujkiewicz2016}.
Bivariate meta-analysis (bvMA) of treatment effects on a surrogate and a final outcome allows for both the validation of a surrogate endpoint and for making predictions of an unobserved treatment effect on the final clinical outcome from observed treatment effects on a surrogate endpoint.

Candidate surrogate endpoints often are not perfect, and the association patterns between the treatment effects on the surrogate and final outcomes may vary between treatments. Whilst bvMA, described in detail in Section \ref{sec.brma}, can be used to model surrogate endpoints, it does not differentiate between the treatment options. This is an important issue when the evidence base consists of multiple trials of different treatments in different populations. Network meta-analysis (NMA) combines data from trials investigating heterogeneous treatment contrasts and has the advantage of estimating effects for all treatment contrasts individually.
At the same time, the consistency assumption allows for combining evidence (and borrowing strength) across the treatment contrasts.
In this paper, we will exploit this framework to model surrogacy relationships by extending previously reported methods of multivariate NMA \cite{achana2014, efthimiou2014, hong2015, jackson2017}.

In the multivariate NMA (mvNMA), true treatment effects on multiple correlated outcomes are assumed to follow separate multivariate distributions for each treatment contrast.
The previously reported mvNMA methods typically simplified the between-studies variance-covariance structures by assuming homogeneity of the correlations and the heterogeneity parameters across the treatment contrasts. We relax this assumption of homogeneity to model in detail different association patterns between the effects on surrogate and final endpoints separately for different treatments or treatment classes.
By allowing such patterns to differ across treatments, the methodology can help to identify treatments for which the surrogacy holds and to improve predictions.
The network consistency assumption provides a framework for combining evidence across treatment contrasts and hence modelling and distinguishing surrogacy patterns within and across treatment contrasts.
These two levels of surrogacy
enable predictions of the treatment effect on the final outcome in a new study investigating either an existing treatment in a new population (within treatment contrast surrogacy) or a new treatment (across treatment surrogacy).
We  extend the second order consistency condition, described by Lu and Ades \cite{lu2009} in a univariate NMA, to the bivariate case, in order to gain additional precision in modelling surrogate relationships.

We illustrate the use of the methods, described in Sections \ref{sec.brma} and \ref{sec.nma}, by fitting them to data from an example in advanced colorectal cancer introduced in Section \ref{sec.ill} (results presented in Section \ref{sec.acrc}).
To illustrate the motivation and the application of the methodology in a more detailed and controlled manner, we apply the methods to simulated data scenarios (Section \ref{sec.ill.sim}).
We fit all models in a Bayesian framework using the OpenBUGS software.

\section{Illustrative example}
\label{sec.ill}
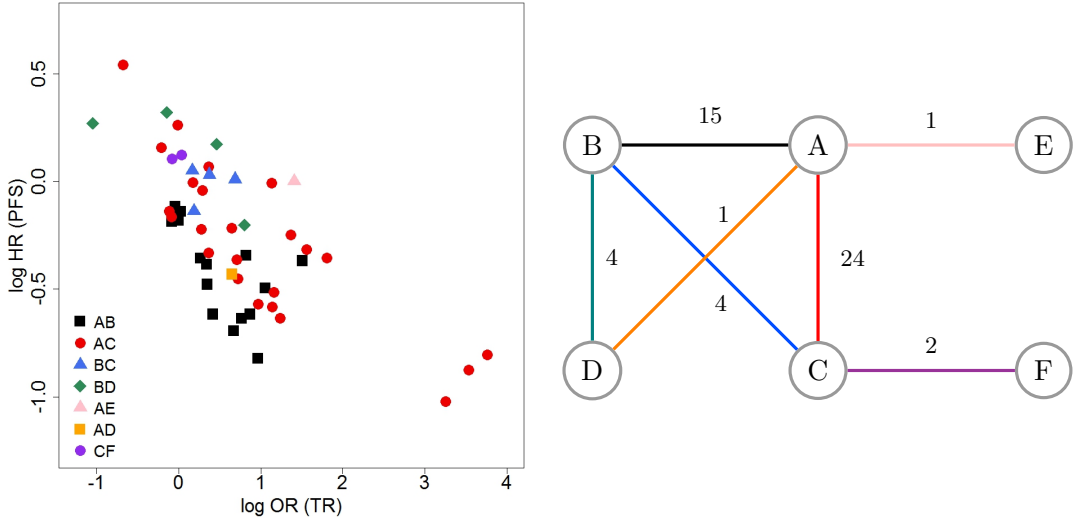
\begin{figure}[h]
\centering
\includegraphics[scale=0.3]{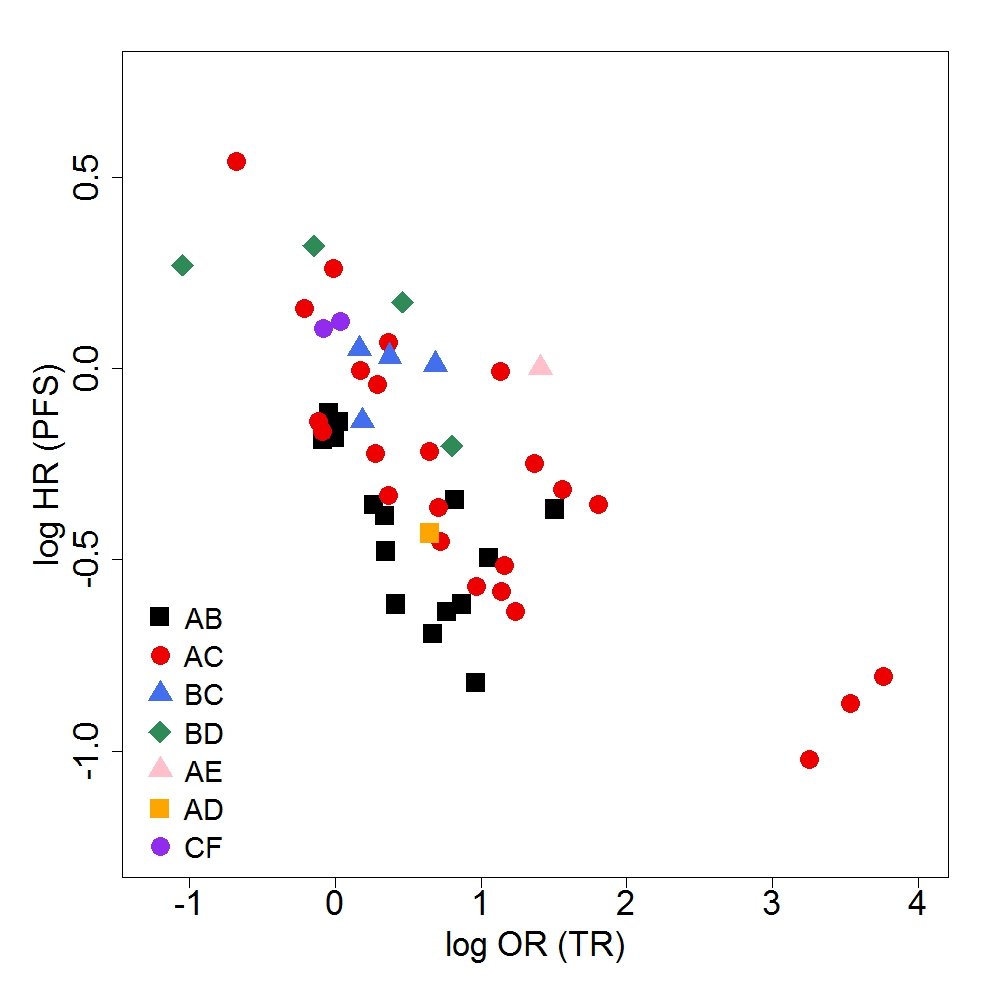}
\begin{tikzpicture}[nodes={draw, circle}, line width=1.2pt]
[inner sep=3mm]
\path node at ( -3,4) [shape=circle,draw=black!40] (b) {B}
node at ( 0,4) [shape=circle,draw=black!40] (a) {A}
node at ( -3,1) [shape=circle,draw=black!40] (d) {D}
node at ( 0,1) [shape=circle,draw=black!40] (c) {C}
node at ( 3,1) [shape=circle,draw=black!40] (f) {F}
node at ( 3,4) [shape=circle,draw=black!40] (e) {E}
node at (0,-1) [shape=circle,draw=white] (ghost) {}
(a) edge[black] node[above, align=flush center, draw=none, xshift= 2pt]
                                {\footnotesize \textcolor[rgb]{0.00,0.00,0.00}{$15$}} (b)
(a) edge[red] node[right, align=flush center, draw=none, xshift= 2pt]
                                {\footnotesize \textcolor[rgb]{0.00,0.00,0.00}{$24$}} (c)
(b) edge[blue!70!cyan] node[below, align=flush center, draw=none, xshift= 6pt, yshift= -8pt]
                                    {\footnotesize \textcolor[rgb]{0.00,0.00,0.00}{$4$}} (c)
(b) edge[teal] node[right, align=flush center, draw=none, xshift= -2pt] {\footnotesize \textcolor[rgb]{0.00,0.00,0.00}{$4$}} (d)
(a) edge[pink] node[above, align=flush center, draw=none] {\footnotesize \textcolor[rgb]{0.00,0.00,0.00}{$1$}} (e)
(a) edge[orange] node[above, align=flush center, draw=none, xshift= 7pt, yshift= 7pt] {\footnotesize \textcolor[rgb]{0.00,0.00,0.00}{$1$}} (d)
(c) edge[violet!80] node[above, align=flush center, draw=none] {\footnotesize \textcolor[rgb]{0.00,0.00,0.00}{$2$}} (f);
\end{tikzpicture}
\caption{Scatter plot and network diagram for the advanced colorectal cancer example, A -- chemotherapy alone, B -- anti-VEGF therapies + chemotherapy,
C -- EGFRi  + chemotherapy, D -- EGFRi + anti-VEGF therapies + chemotherapy, E -- anti-IGF1R , F -- anti-IgG2 + chemotherapy}
\label{fig_acrc_scatter}
\end{figure}

We use data from randomised controlled trials (RCTs) in advanced colorectal cancer (aCRC), investigating a range of different treatment options, to illustrate how the hierarchical bvNMA models, proposed in this paper, can differentiate the association patterns between the treatment contrasts or classes whilst borrowing strength across treatment contrasts.
Data were obtained from four published systematic reviews of RCTs investigating pharmacological treatments in aCRC, categorised into classes with respect to their mechanism of action. These were targeted therapies including antiangiogenic treatments targeting vascular endothelial growth factor (anti-VEGF) \cite{wagner2009}, anti epidermal growth factor receptor inhibitors (EGFRi) \cite{chan2017}, humanised monoclonal antibody targeting integrin receptors (anti-IgG2) and monoclonal antibody targeting the type 1 insulin-like growth factor receptor (anti-IGF1R) \cite{mocellin2017} or chemotherapies compared to the targeted therapies \cite{mocellin2017, kumachev2015} and combinations of these therapies. 15 RCTs investigated use of anti-VEGF with chemotherapy vs.  chemotherapy alone, 24 RCTs of EGFRi with chemotherapy vs.  chemotherapy alone, 4 RCTs of EGFRi with chemotherapy vs.  anti-VEGF with chemotherapy, 4 RCTs of EGFRi with anti-VEGF and chemotherapy vs.  anti-VEGF with chemotherapy, one study (in two subgroups of population) of anti-IgG2 with EGFRi  and chemotherapy vs. EGFRi with chemotherapy, one study of anti-IGF1R with chemotherapy vs. chemotherapy alone and one of EGFRi with anti-VEGF and chemotherapy vs. chemotherapy alone.
The treatments are summarised in the network diagram of Figure \ref{fig_acrc_scatter}.
Tumour response (TR) is used as an example of a potential surrogate endpoint to progression free survival (PFS).
The models are applied to data representing treatment effect on the two outcomes on log odds ratio (OR) scale for TR and log hazard ratio (HR) scale for PFS. The scatter plot in Figure \ref{fig_acrc_scatter} illustrates the association patterns between the treatment effects on the two outcomes for each treatment contrast.

%%%%%%%%%%%%%%%%%%%%%%%%%%%%%%%%%%%%%%%%%%%%%%%%%%%%%%%%%%%%%%%%%%%%%%%%%%%%%%%%%%%%%%%%%%%%%%%%%%%%%%%%%%%%%%%%%%%%%%%%%%%%%%%%%%%%%%%%%%%%%%%%%%%%%%%%%%%%%%%%%%%%%%%%%%%%%%%%%%%
\section{Bivariate random effects meta-analysis (BRMA)}
\label{sec.brma}
The bivariate random effects meta-analysis (BRMA) model for correlated and normally distributed treatment effects on two outcomes $Y_{1i}$ and $Y_{2i}$ is usually presented in the form described by
van Houwelingen \emph{et al.} \cite{vanhouwelingen2002} and
Riley \emph{et al.}
\cite{riley2007}:
\begin{equation}
\left(
\begin{array}{c}
Y_{1i}\\
Y_{2i}\\
\end{array}
\right) \sim \rm{N}
\left\{
\left(
\begin{array}{c}
\mu_{1i}\\
\mu_{2i}\\
\end{array}\right), \mathbf{\Sigma_i}
\right\}, \;
\mathbf{\Sigma_i}=\left(
\begin{array}{cc}
\sigma_{1i}^2 & \sigma_{1i}\sigma_{2i}\rho_{wi}\\
\sigma_{1i}\sigma_{2i}\rho_{wi} & \sigma_{2i}^2
\end{array}
\right)
\label{brma-w}
\end{equation}
\begin{equation}
\left(
\begin{array}{c}
\mu_{1i}\\
\mu_{2i}\\
\end{array}
\right) \sim \rm{N}
\left\{
\left(
\begin{array}{c}
\beta_{1}\\
\beta_{2}\\
\end{array}\right), \mathbf{T}
\right\}, \;
\mathbf{T}=\left(
\begin{array}{cc}
\tau_{1}^2 & \tau_{1}\tau_{2}\rho\\
\tau_{1}\tau_{2}\rho & \tau_{2}^2
\end{array}
\right).\\
\label{brma-b}
\end{equation}
In this model, the treatment effects on the surrogate endpoint $Y_{1i}$ and  on the final outcome $Y_{2i}$ are
assumed to estimate the correlated true effects $\mu_{1i}$ and $\mu_{2i}$ with corresponding within-study variances
$\sigma_{1i}^2$ and $\sigma_{2i}^2$  of the estimates and the within-study correlation $\rho_{wi}$
between them.
In this hierarchical framework, these true study-level effects follow a bivariate normal distribution with means $\left(\beta_1, \beta_2\right)$ corresponding to the two outcomes, between-studies
variances $\tau_{1}^2$ and $\tau_{2}^2$  and a between-studies correlation $\rho$.
Equation (\ref{brma-w}) represents the within-study model and (\ref{brma-b}) is the between-studies model.
To implement the model in the Bayesian framework, prior distributions are placed on the mean effects, for example vague prior distributions $\beta_{1}\sim N(0,10^4)$, $\beta_{2}\sim N(0,10^4)$ and on the between-studies variances and correlation.

In the general case, for any number of outcomes, a prior distribution has to be placed on the whole variance-covariance matrix or the correlation matrix in such a way to ensure that the variance-covariance matrix is positive semi-definite. This can be achieved by placing an inverse Wishart prior distribution on the variance-covariance matrix \cite{wei2013} or by using a separation strategy, with spherical \cite{lu2009, wei2013} or Cholesky decomposition \cite{wei2013} of the correlation matrix. Alternatively, a product normal formulation of the between-studies model can be used where the model is parameterised in the form of a series of univariate conditional distributions, ensuring that the relationships between the parameters of the model (regression coefficients and conditional variances) and the elements of the between-studies variance-covariance matrix result in the positive semi-definite between-studies variance-covariance matrix \cite{bujkiewicz2013, bujkiewicz2015, bujkiewicz2016}.

In the bivariate case, such as considered here,  positive semi-definiteness can be achieved by placing prior distributions on the variances, which are restricted to plausible positive values, for example by placing uniform prior distributions on the corresponding standard deviations $\tau_j \sim U(0,2)$, and by choosing a prior distribution for the correlation which restricts it to values between $-1$ and $+1$. Here we use a beta distribution to construct such a prior: $\frac{\rho+1}{2}\sim Beta(1.5,1.5)$, as in
Burke \emph{et al.} \cite{burke2016}, with details also in the Appendix \ref{appendix2}.

%%%%%%%%%%%%%%%%%%%%%%%%%%%%%%%%%%%%%%%%%%%%%%%%%%%%%%%%%%%%%%%%%%%%%%%%%%%%%%%%%%%%%%%%%%%%%%%%%%%%%%%%%%%%%%%%%%%%%%%%%%%%%%%%%%%%%%%%%%%%%%%%%%%%%%%%%%%%%%%%%%%%%%%%%%%%%%%%%%%
\section{Bivariate network meta-analysis (bvNMA)}
\label{sec.nma}
The BRMA model is typically used to obtain average effects on two correlated outcomes,
 %while taking into account the correlation between them,
 for studies of the same treatment or treatment class.
In the context of surrogate endpoints, data on a range of treatments
are typically used.
BRMA assumes exchangeability of the treatment effects from all studies regardless of the treatment contrast, by assuming that the true effects follow a common (here bivariate normal) distribution (as in Eq. (\ref{brma-b})).
This model works well for strong surrogate relationships across all treatments.
However, the association pattern between treatment effects measured on the surrogate and final outcomes may depend on the treatment contrast. Network meta-analysis can differentiate between treatment contrasts and in bivariate form can be applied to modelling such surrogate relationships within and across treatment contrasts.
We first discuss general bvNMA models in Section \ref{bvNMAgen} which describe the within-treatment surrogate relationships in the network. In Section \ref{bvNMAex} we extend these models to allow for an additional level of surrogacy across treatment contrasts.
\subsection{General bvNMA models}
\label{bvNMAgen}
\subsubsection{Model 1a: bvNMA}
\label{sec.nma.het}
To model correlated treatment effects on a surrogate endpoint and a final outcome for which the correlation varies according to the treatment contrasts, the assumption made by BRMA, that the true effects follow a common distribution, can be replaced by an assumption that the true effects corresponding to different treatment contrasts follow separate distributions.
This naturally leads to relaxing the assumption of homogeneity of the between-studies covariance matrix, allowing its elements (the between-studies correlations $\rho_{1kl,2kl}$ between the treatment effects $l$ vs. $k$ on the surrogate (1) and final (2) outcomes  and the heterogeneity parameters for the treatment effects on the two outcomes $\tau_{1kl}^2$ and $\tau_{2kl}^2$) to vary across the treatment contrasts $kl$.

To take into account the network structure of the data, we model the treatment effect differences $Y_{jkli}$ between treatments $k$ and $l$ in study $i$ for outcome $j=1,2$, as follows:
\begin{equation}
\left(
\begin{array}{c}
Y_{1kli}\\
Y_{2kli}\\
\end{array}
\right) \sim \rm{N}
\left\{
\left(
\begin{array}{c}
\mu_{1kli}\\
\mu_{2kli}\\
\end{array}\right), \mathbf{\Sigma_i}
\right\}, \;
\mathbf{\Sigma_i}=\left(
\begin{array}{cc}
\sigma_{1kli}^2 & \sigma_{1kli}\sigma_{2kli}\rho_{wkli}\\
\sigma_{1kli}\sigma_{2kli}\rho_{wkli} & \sigma_{2kli}^2
\end{array}
\right)
\label{eq-bvnma-w-het}
\end{equation}
\begin{equation}
\left(
\begin{array}{c}
\mu_{1kli}\\
\mu_{2kli}\\
\end{array}
\right) \sim \rm{N}
\left\{
\left(
\begin{array}{c}
%d_{1,1k} - d_{1,1b}\\
%d_{2,1k} - d_{2,1b}\\
d_{1kl}\\
d_{2kl}\\
\end{array}\right),
%\mathbf{T}
%\right),
\;
%\mathbf{T}=
\left(
\begin{array}{cc}
\tau_{1kl}^2 & \tau_{1kl}\tau_{2kl}\rho_{1kl,2kl}\\
\tau_{1kl}\tau_{2kl}\rho_{1kl,2kl} & \tau_{2kl}^2
\end{array}
\right)
\right\}
\label{eq-bvnma-b-het}
%\nonumber
\end{equation}
where $k$ and $l$ denote baseline (control) and experimental treatments respectively in a study $i$, $\mu_{jkli}$ denote the random true treatment effects (differences between the effects of treatments $k$ and $l$) on outcome $j$ in study $i$ and, and the $d_{jkl}$ are mean treatment effect differences between treatments $k$ and $l$ for each outcome $j$.
We use the first-order consistency assumption, as described by
Lu and Ades \cite{lu2009}, extended here to the bivariate case. For any three treatments $(b, k, l)$, the treatment differences $\mathbf(\mu_{jkli})$ satisfy the following transitivity relations
\begin{equation}
\left(
\begin{array}{c}
\mu_{1kli}\\
\mu_{2kli}\\
\end{array}
\right)
=
\left(
\begin{array}{c}
\mu_{1bli} - \mu_{1bki}\\
\mu_{2bli} - \mu_{2bki}\\
\end{array}
\right).
\label{eq-trans}
\end{equation}
Taking the expectation on both sides gives the consistency equations for the first-order moments
\begin{equation}
\left(
\begin{array}{c}
d_{1kl}\\
d_{2kl}\\
\end{array}
\right)
=
\left(
\begin{array}{c}
d_{1bl} - d_{1bk}\\
d_{2bl} - d_{2bk}\\
\end{array}
\right)
\label{eq-first-ord-consis}
\end{equation}
which represent the relationships between the treatment contrasts in the population.
When $b=1$ is a common reference treatment in the network, the treatment effects of each treatment $k$ in the network relative to this common reference treatment $1$; the $d_{j,1k}$ are referred to as basic parameters for each outcome $j$, with $d_{j,11}=0$ and the others are given prior distributions:
\begin{equation}
d_{j,1k}  \sim   N(0, 10^3).
\label{eq-prior-d}
\end{equation}
Prior distributions are also placed on the elements of the between-studies variance-covariance matrix.
Similarly as in BRMA, prior distributions for the heterogeneity parameters are selected in a way to ensure that they are restricted to plausible positive values, such as $\tau_{jkl}\sim unif(0,2)$ and for the correlations to ensure restriction to the values between $-1$ and $1$, e.g.  $\frac{\rho_{kl}+1}{2}\sim Beta(1.5,1.5)$, thus guaranteeing that the variance-covariance matrix is positive semi-definite.

%%%%%%%%%%%%%%%%%%%%%%%%%%%%%%%%%%%%%%%%%%%%%%%%%%%%%%%%%%%%%%%%%%%%%%%%%%%%%%%%%%%%%%%%%%%%%%%%%%%%%%%%%%%%%%%%%%%%%%%%%%%%%%%%%%%%%%%%%%%%%%%%%%%%%%%%%%%%%%%%%%%%%%%%%%%%%%%%%%%
\subsubsection{Model 1b: bvNMA with second order consistency}
\label{sec.nma.het.2or}
The above meta-analytic model 1a assumes consistency of treatment effects on both outcomes.
This implies some constraints on the between-studies variance-covariance matrices which can be explicitly introduced to the bvNMA model (\ref{eq-bvnma-w-het})--(\ref{eq-bvnma-b-het}) by  assuming the consistency of the second-order moments.
To do so, we extend the approach proposed by
Lu and Ades \cite{lu2009} to the bivariate case by taking variances on both sides of the transitivity equation (\ref{eq-trans}), which gives
\begin{eqnarray}
\begin{array}{c}
\left(
\begin{array}{cc}
\tau_{1kl}^2 & \tau_{1kl} \tau_{2kl} \rho_{1kl,2kl}\\
\tau_{1kl} \tau_{2kl} \rho_{1kl,2kl} & \tau_{2kl}^2 \\
\end{array}
\right)
\vspace{0.5cm}
\\
=
\left(
\begin{array}{cc}
var(\mu_{1bli} - \mu_{1bki}) & cov(\mu_{1bli} - \mu_{1bki}, \mu_{2bli} - \mu_{2bki}) \\
cov(\mu_{1bli} - \mu_{1bki},\mu_{2bli} - \mu_{2bki})  & var(\mu_{2bli} - \mu_{2bki})\\
\end{array}
\right)
\end{array}
\end{eqnarray}
leading to  the following relationship between the variances for any three treatments $(b, k, l)$ and for both outcomes $j=1,2$:
\begin{eqnarray}
\tau_{jkl}^2  =  \tau_{jbk}^2+\tau_{jbl}^2 - 2 \rho_{jbk,jbl} \tau_{jbk} \tau_{jbl} \le (\tau_{jbk}+\tau_{jbl})^2,
%\label{eq-sec-ord-consist}
\end{eqnarray}
which gives the second-order consistency conditions (triangle inequalities):
\begin{equation}
\vert \tau_{jbl} - \tau_{jbk} \vert \le \tau_{jkl} \le \tau_{jbl} + \tau_{jbk}.
\label{eq-sec-ord-consist}
\end{equation}
In addition, the following condition applies to the covariances:
\begin{equation}
\tau_{1kl} \tau_{2kl} \rho_{1kl,2kl}  =  \tau_{1bl}\tau_{2bl}\rho_{1bl,2bl} + \tau_{1bk}\tau_{2bk}\rho_{1bk,2bk}
                    - \tau_{1bl}\tau_{2bk}\rho_{1bl,2bk} - \tau_{1bk}\tau_{2bl}\rho_{1bk,2bl},
\label{eq-sec-ord-consist2}
\end{equation}
which implies further constraints that are more complex than those in Eq. (\ref{eq-sec-ord-consist}).

To ensure that prior distributions for heterogeneous variance-covariance matrices are appropriate, i.e. to maintain the second-order consistency condition for any three treatments in the network, bivariate ancillary parameters are used, extending the univariate model by

To ensure that prior distributions for heterogeneous variance-covariance matrices are appropriate, i.e. to maintain the second-order consistency condition for any three treatments in the network, bivariate ancillary parameters are used, similarly as by Lu and Ades \cite{lu2009}, allowing the between-studies variance-covariance matrices to be represented as
\begin{eqnarray}
& \left(
\begin{array}{cc}
\tau_{1kl}^2 & \tau_{1kl} \tau_{2kl} \rho_{1kl,2kl}\\
\tau_{1kl} \tau_{2kl} \rho_{1kl,2kl} & \tau_{2kl}^2 \\
\end{array}
\right)
= & \nonumber\\
& \; & \nonumber\\
& \left(
\begin{array}{cc}
\gamma_{1k}^2+\gamma_{1l}^2 - 2 \xi_{1k,1l} \gamma_{1k} \gamma_{1l}
&
\begin{array}{c}
\gamma_{1k}\gamma_{2k}\xi_{1k,2k} - \gamma_{1k}\gamma_{2l}\xi_{1k,2l}\\
-\gamma_{1l}\gamma_{2k}\xi_{1l,2k} + \gamma_{1l}\gamma_{2l}\xi_{1l,2l}
\end{array}
\\
\; \\
\begin{array}{c}
\gamma_{1k}\gamma_{2k}\xi_{1k,2k} - \gamma_{1k}\gamma_{2l}\xi_{1k,2l}\\
-\gamma_{1l}\gamma_{2k}\xi_{1l,2k} + \gamma_{1l}\gamma_{2l}\xi_{1l,2l}
\end{array}
&
\gamma_{2k}^2+\gamma_{2l}^2 - 2 \xi_{2k,2l} \gamma_{2k} \gamma_{2l}
\end{array}
\right)&
\label{eq-sec-ord-consist-ancillary}
\end{eqnarray}
where
$\gamma_{jk}^2$ and  $\gamma_{jl}^2$ are the ancillary parameters: 
 the variances of two correlated random effects $\zeta_{jki}$ and $\zeta_{jli}$ corresponding  to treatment arms $k$ and $l$ (for each outcome $j=1,2$).
Prior distributions for the set of between-studies standard deviations $\tau_{jkl}$ for each outcome $j$ and each pair of treatments $k$ and $l$ can be given by constructing a prior distribution for a covariance matrix $\Gamma$ composed of the standard deviations $\gamma_{jk}$ and the correlations $\xi_{jk,j'l}$ between the effects $\zeta_{jki}$ and $\zeta_{j'li}$, for $j, j'=1,2$ and $k,l = 1, \dots, n_{t}$, where $n_t$ is the number of treatments in the network.
For the set of values of the elements of matrix $\Gamma$ to give a resulting set of standard deviations $\tau_{jkl}$ that satisfy the second-order consistency rules (\ref{eq-sec-ord-consist}) and (\ref{eq-sec-ord-consist2}), the matrix $\Gamma$ has to be positive semi-definite.
This is achieved using a separation strategy with a Cholesky decomposition:  $\Gamma=V^{1/2} R V^{1/2}$, where $V^{1/2}$ is a $2 n_t \times 2 n_t$ diagonal matrix of the standard deviations $\gamma_{11}, \gamma_{21}, \dots, \gamma_{1n_t}, \gamma_{2n_t}$ and $R$ is a positive semi-definite $2 n_t \times 2 n_t$ matrix of correlations $\xi_{jk,\hat{j}l}$ (block matrix consisting of $n_t\times n_t$ blocks that are of $2\times 2$ dimension).
Matrix $R$ is represented as $R=L^T L$ with $L$ being a $2 n_t \times 2 n_t$  upper triangular matrix.
To obtain the elements of the matrix $L$, we extended the method  by Wei and Higgins (2013) who describe it for a four dimensional matrix case.
Prior distributions are placed on the standard deviations, which need to be restricted to positive values, for example $\gamma_{j,k} \sim unif(0,2)$.
The prior distributions placed on the ancillary variances and correlations give implied prior distributions on the between-studies correlations and standard deviations through the formulae (\ref{eq-sec-ord-consist-ancillary}).

%%%%%%%%%%%%%%%%%%%%%%%%%%%%%%%%%%%%%%%%%%%%%%%%%%%%%%%%%%%%%%%%%%%%%%%%%%%%%%%%%%%%%%%%%%%%%%%%%%%%%%%%%%%%%%%%%%%%%%%%%%%%%%%%%%%%%%%%%%%%%%%%%%%%%%%%%%%%%%%%%%%%%%%%%%%%%%%%%%%
\subsubsection{Model 1c: bvNMA with second order consistency and similarity of the ancillary parameters}
\label{sec.nma.het.2or.ex}
To ensure second order consistency of the treatment effects in the network, model 1b requires estimation of relatively many parameters. Where data for each treatment $k$ is only available from a limited number of studies, it may be difficult to estimate the individual variances $\gamma_{1k}^2$.
To overcome this issue, exchangeability of the ancillary variances may be assumed by replacing individual prior distributions for these parameters with a common distribution;
\begin{equation}
\label{eq-var-sim}
\gamma_{jk}^2  \sim  N(0,v_j)I(0,) \;\;\; {\rm and} \; \;\; v_j  \sim  \Gamma(1.0,0.01).
\end{equation}

%%%%%%%%%%%%%%%%%%%%%%%%%%%%%%%%%%%%%%%%%%%%%%%%%%%%%%%%%%%%%%%%%%%%%%%%%%%%%%%%%%%%%%%%%%%%%%%%%%%%%%%%%%%%%%%%%%%%%%%%%%%%%%%%%%%%%%%%%%%%%%%%%%%%%%%%%%%%%%%%%%%%%%%%%%%%%%%%%%%
\subsubsection{Model 1d: bvNMA assuming homogeneity of the between-studies variance-covariance matrix}
\label{sec.nma.hom}
When networks are sparse, instead of assuming similarity between the heterogeneity parameters, as in Eq. (\ref{eq-var-sim}), homogeneity of the between-studies variance-covariance matrix can be assumed, which is a common assumption in the multivariate network meta-analysis or the univariate network meta-analysis of multi-arm trials \cite{lu2004}.
The between-studies equation (\ref{eq-bvnma-b-het}) then becomes
\begin{equation}
\left(
\begin{array}{c}
\mu_{1kli}\\
\mu_{2kli}\\
\end{array}
\right) \sim \rm{N}
\left\{
\left(
\begin{array}{c}
d_{1,1l} - d_{1,1k}\\
d_{2,1l} - d_{2,1k}\\
\end{array}\right),
%\mathbf{T}
%\right),
\;
\mathbf{T}=
\left(
%\begin{array}{cc}
%\tau_{1kl}^2 & \tau_{1kl}\tau_{2kl}\rho_{kl}\\
%\tau_{1kl}\tau_{2kl} & \tau_{2kl}^2
%\end{array}
\begin{array}{cc}
\tau_{1}^2 & \tau_{1}\tau_{2}\rho\\
\tau_{1}\tau_{2}\rho & \tau_{2}^2
\end{array}
\right)
\right\}
\label{eq-bvnma-b-hom}
%\nonumber
\end{equation}
where $\mu_{jkli}$ is the random treatment effect difference between treatments $k$ and $l$ on outcome $j$ in study $i$ and $d_{j,1k}$ is the mean treatment effect difference between treatment $k$ and the reference treatment $1$ for outcome $j$, with $d_{j,11}=0$. A prior distribution is placed on each basic parameter, $d_{j,1k}  \sim   N(0, 10^3)$.
A prior distribution is also placed on the elements of the common between-studies variance-covariance matrix: the between-studies standard deviations, $\tau_j \sim Unif(0,2)$ and the correlation, $\frac{\rho +1}{2} \sim Beta(1.5,1.5)$.

\subsubsection{Surrogacy criteria:}
\label{sec.sur.cri1}
When using the bivariate NMA models 1a-c to evaluate surrogate endpoints within treatment contrast $kl$, perfect surrogacy means that
\begin{equation}
\rho_{1kl,2kl}=\pm 1 \;\; {\rm and} \;\; \mu_{1kli}=0 \; \Leftrightarrow \; \mu_{2kli}=0.
\label{sur.crit.1a}
\end{equation}
This condition is equivalent to assuming that in the linear relationship between true treatment effects on the final and surrogate endpoints, the intercept is zero, $\mu_{2kli}=const \times \mu_{1kli}$ and the conditional variance (of the treatment effect on the final outcome conditional on the effect on the surrogate endpoint) is also zero, as in the standard surrogacy models \cite{daniels1997,bujkiewicz2015} (for more details see Appendix \ref{sec.sur.crit.apx}).

This criterion describes the between-studies (and populations) surrogacy relationships within the treatment contrasts.
The network structure, with the unique surrogate relationship for each treatment contrast, can help us to disentangle information about a surrogate relationship for a particular treatment and to make better predictions, in particular when these relationships are clearly distinct, as illustrated in Section \ref{sec.ill.sim}.
These individual surrogacy relationships enable predicting the treatment effect on the final outcome from the treatment effect measured on a surrogate endpoint in a new study investigating an existing treatment  either in a new population or with a different (but already present in the network) control treatment.

When evaluating surrogate endpoints with model 1d, similarly as for models 1a-c the random effects are assumed to follow separate distributions for studies evaluating different treatment contrasts $kl$ but the between-studies correlations and heterogeneity parameters are assumed to be the same across the treatment contrasts.
In this case, the perfect surrogacy means that
\begin{equation}
\rho=\pm 1 \;\; {\rm and} \;\; \mu_{1kli}=0 \; \Leftrightarrow \; \mu_{2kli}=0
\label{sur.crit.1b}
\end{equation}
which differs from the criteria for the models 1a-c  with respect to the common correlation across the treatment contrasts.
The strength of the association between the treatment effects on the surrogate endpoint and the final clinical outcome will be the same across the treatment contrasts when using model 1d, similarly as when using a pairwise meta-analysis such as BRMA.
However, assuming that the random effects for different treatment contrasts follow separate distributions in the bvNMA, we model these association patterns in more detail compared to when using BRMA. In addition, taking into account the network structure of the data results in borrowing of strength across the treatment contrasts when evaluating surrogate relationships.

%%%%%%%%%%%%%%%%%%%%%%%%%%%%%%%%%%%%%%%%%%%%%%%%%%%%%%%%%%%%%%%%%%%%%%%%%%%%%%%%%%%%%%%%%%%%%%%%%%%%%%%%%%%%%%%%%%%%%%%%%%%%%%%%%%%%%%%%%%%%%%%%%%%%%%%%%%%%%%%%%%%%%%%%%%%%%%%%%%%
\subsection{bvNMA models with borrowing of strength across treatment contrasts}
\label{bvNMAex}
\begin{figure}[h]
%\hspace{1cm}
\centering
\begin{tikzpicture}[nodes={draw, circle}]
\centering
[inner sep=3mm]
\path node at ( 0,1.6) [shape=circle] (f) {A}
node at ( -1,0) [shape=circle] (s1) {B}
node at ( 1,0) [shape=circle] (s2) {C}
node at ( 2,1.6) [shape=circle] (s3) {D}
(s1) edge  (f)
(s2) edge (f)
(s1) edge (s2)
(s2) edge node[pos=0.3, right, draw=none] {\footnotesize \textcolor[rgb]{0.00,0.00,0.00}{1}} (s3);
\end{tikzpicture}
\hspace{1cm}
\begin{tikzpicture}[nodes={draw, circle}
%, line width=2pt
]
\centering
[inner sep=3mm]
\path node at ( 0,1.6) [shape=circle] (f) {A}
node at ( -1,0) [shape=circle] (s1) {B}
node at ( 1,0) [shape=circle] (s2) {C}
(s1) edge  (f)
(s2) edge (f)
(s1) edge (s2);
\end{tikzpicture}
\caption{Example network diagram: data on  effect of new  treatment D only available in one new study and only measured on surrogate endpoint (left) but not on the final outcome (right).\label{fig1}}
\end{figure}
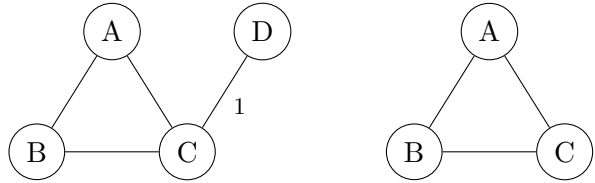

Models 1a--d describe the relationships between the treatment effects on correlated outcomes when for each treatment contrast in the network there is at least one study reporting the treatment effects on both outcomes.
In the situation where we want to predict the treatment effect on the final outcome from the effect on the surrogate endpoint in a new study evaluating a new treatment,  there will be only this one study reporting the effect for that treatment and only on the surrogate endpoint.
In this case the above models 1a--d will result in an estimate of the average effect on the final outcome for the new treatment based solely on the prior distribution,  because no data are available for the effect of this treatment on the final outcome. For example, as in the network depicted in Figure  \ref{fig1}, if we select treatment A as the reference treatment 1, then the average treatment effect for the new treatment $D$ on the final outcome equals $d_{2, CD}=d_{2,AD}-d_{2,AC}$ and the estimate of the basic parameter $d_{2,AD}$ would be based on the prior distribution only. As a result, the estimate of the between-studies correlation $\rho_{1CD,2CD}$ will be based solely on the prior distribution. Therefore the predicted treatment effect on the final outcome from the treatment effect on the surrogate endpoint for the new treatment will not be meaningful.
The models 1a--d are designed to evaluate surrogacy patterns across different populations within a treatment contrast and make predictions of a treatment effect on the final outcome in a new study investigating an existing treatment only, but perhaps in a new population.

\subsubsection{Model 2a: assuming exchangeability of treatments}
\label{sec.nma.het.ex}
To overcome the above issue, an exchangeability assumption can be made to
allow for the relationships between the average effects on the two outcomes to be similar across treatment options.
To achieve this for model 1a (Eqs (\ref{eq-bvnma-w-het}), (\ref{eq-bvnma-b-het}) and (\ref{eq-first-ord-consis})), instead of placing a prior distribution on each basic parameter (Eq. (\ref{eq-prior-d})), it is assumed that the pooled effects for each treatment arm $k$, $\theta_{jk}$ on the two outcomes $j=1,2$, are exchangeable and correlated:
\begin{equation}
\label{eq-exchange-1}
d_{j1k}  =  \theta_{jk}-\theta_{j1},
\end{equation}
\begin{equation}
\label{eq-exchange-2}
\left(
\begin{array}{c}
\theta_{1k} \\
\theta_{2k} \\
\end{array}
\right)
 \sim
N
\left(
\left(
\begin{array}{c}
\eta_1 \\ \eta_2 \\
\end{array}
\right),
\frac{1}{2}
\left(
\begin{array}{cc}
\omega_1^2 & \omega_1 \omega_2 \rho_t \\
\omega_1 \omega_2 \rho_t & \omega_2^2 \\
\end{array}
\right)
\right)
\end{equation}
for $k=1,\dots, n_t$.
Assuming exchangeability of the effects in each treatment arm $\theta_{jk}$, rather than of the basic parameters $d_{j1k}$ which are the effects relative to a common reference treatment, ensures that the prior distributions for each pair $(\theta_{1k},\theta_{2k})$ are independent.

The modelled data $Y_{jkli}$ are the treatment effect differences, thus the ancillary parameters $\theta_{jk}$  are not identifiable. However, we are not interested in estimating these parameters, but in estimation of the association between the average treatment effect differences on the two outcomes: $d_{jkl}$, $j=1,2$.
The above formulae imply the association between the average effects:
\begin{equation}
\label{eq-exchange-3b}
\left(
\begin{array}{c}
d_{1kl} \\
d_{2kl} \\
\end{array}
\right)
 \sim
N
\left\{
\left(
\begin{array}{c}
0 \\ 0 \\
\end{array}
\right),
\left(
\begin{array}{cc}
\omega_1^2 & \omega_1 \omega_2 \rho_t \\
\omega_1 \omega_2 \rho_t & \omega_2^2 \\
\end{array}
\right)
\right\},
\end{equation}
$k=2,\dots, n_t$,
cancelling out the mean values $\eta_j$.
Therefore the mean values can be set arbitrarily to a constant, for example $\eta_j=0$.
Prior distributions are placed on the elements of the covariance matrix: $\omega_j \sim Unif(0,2)$ and $\frac{\rho_t +1}{2} \sim Beta(1.5,1.5)$.
Borrowing of strength across treatments by relating these parameters enables the model to predict the effect of a new treatment on the final outcome, as in the network scenario in Figure \ref{fig1}.

%%%%%%%%%%%%%%%%%%%%%%%%%%%%%%%%%%%%%%%%%%%%%%%%%%%%%%%%%%%%%%%%%%%%%%%%%%%%%%%%%%%%%%%%%%%%%%%%%%%%%%%%%%%%%%%%%%%%%%%%%%%%%%%%%%%%%%%%%%%%%%%%%%%%%%%%%%%%%%%%%%%%%%%%%%%%%%%%%%%
\subsubsection{Model 2b: assuming exchangeability of treatments, with second order consistency constraints}
\label{sec.nma.het.ex.no2or}
We  extend  model 1b, described by the general mvNMA  (\ref{eq-bvnma-w-het})--(\ref{eq-bvnma-b-het}) along with the first (\ref{eq-first-ord-consis}) and second order  (\ref{eq-sec-ord-consist}--\ref{eq-sec-ord-consist2}) consistency assumptions, to allow for the  exchangeability of basic parameters. Similarly as in model 2a, this replaces the individual prior distributions on the basic parameters (\ref{eq-prior-d}) with the formulae (\ref{eq-exchange-1})--(\ref{eq-exchange-2}), with prior distributions as in Section \ref{sec.nma.het.ex}.

%%%%%%%%%%%%%%%%%%%%%%%%%%%%%%%%%%%%%%%%%%%%%%%%%%%%%%%%%%%%%%%%%%%%%%%%%%%%%%%%%%%%%%%%%%%%%%%%%%%%%%%%%%%%%%%%%%%%%%%%%%%%%%%%%%%%%%%%%%%%%%%%%%%%%%%%%%%%%%%%%%%%%%%%%%%%%%%%%%%
\subsubsection{Model 2c: assuming exchangeability of treatments, with second order consistency constraints and exchangeable variances in treatment arms}
This model is an extension of model 1c, assuming similarity across the ancillary parameters described by (\ref{eq-var-sim}), to allow for the  exchangeability of the basic parameters using the formulae (\ref{eq-exchange-1})--(\ref{eq-exchange-2}), as in models 2a and 2b.

%%%%%%%%%%%%%%%%%%%%%%%%%%%%%%%%%%%%%%%%%%%%%%%%%%%%%%%%%%%%%%%%%%%%%%%%%%%%%%%%%%%%%%%%%%%%%%%%%%%%%%%%%%%%%%%%%%%%%%%%%%%%%%%%%%%%%%%%%%%%%%%%%%%%%%%%%%%%%%%%%%%%%%%%%%%%%%%%%%%
\subsubsection{Model 2d: assuming exchangeability of treatments and homogeneity of the between-studies variance-covariance matrix}
\label{sec.nma.hom.ex}
To extend model 1d to allow for the exchangeability of the correlated basic parameters, we used equations (\ref{eq-bvnma-w-het}) and (\ref{eq-bvnma-b-hom}) together with the exchangeability model (\ref{eq-exchange-1})--(\ref{eq-exchange-2}) on the basic parameters.
As in model 1d, prior distributions are placed on the elements of the common between-studies variance-covariance matrix, the between-studies standard deviations, $\tau_j \sim Unif(0,2)$ and the correlation,  $\frac{\rho + 1}{2} \sim Beta(1.5,1.5)$.

\subsubsection{Surrogacy criteria}
\label{sec.sur.cri2}
Similarly as for models 1a-d, the surrogate relationship between the treatment effects on the surrogate endpoint and the final clinical outcome is perfect when the correlation $\rho_{kl}=\pm 1$ ($\rho\pm 1$ for model 2d) and no treatment effect on the surrogate endpoint ($\mu_{1kli}=0$) will imply no effect on the final outcome ($\mu_{2kli}=0$) in the same study $i$, as described by the formulae (\ref{sur.crit.1a}) and (\ref{sur.crit.1b}).
This is the surrogacy relationship between studies (or populations) described within each treatment contrast $kl$ and it is assumed to differ across treatment contrasts.
In addition to this, another level of surrogate relationship is described by models 2a-d, the across-treatments surrogacy.
Such a surrogate relationship would be perfect if \begin{equation}
\rho_{t}=\pm 1, \; {\rm and} \; d_{1,kl}=0 \; \Leftrightarrow \; d_{2,kl}=0,
\end{equation}
a zero average treatment effect difference between treatments $k$ and $l$ on the surrogate endpoint will imply zero average treatment effect also on the final clinical endpoint for the same treatment contrast.
%a zero treatment effect of a given treatment compared to a common reference treatment on the surrogate endpoint will imply zero average effect of that treatment compared to the reference treatment also on the final clinical endpoint.
This surrogacy relationship allows prediction of the treatment effect on the final outcome from the treatment effect measured on a surrogate endpoint in a new study investigating a new treatment.
\subsection{Summary and discussion of models}
Bivariate models for network meta-analysis described in this Section differ in their specific assumptions about the heterogeneity between studies and treatments.
All components of the between-study models for all models, 1a--d and 2a--d, are summarised in Table \ref{tab-models}.
The models allow the correlations between the treatment effects on the surrogate and  final endpoints to vary to a different degree, and which model is applied in practice can be determined based on the available evidence.
The models reduce to the standard meta-analysis model for surrogate endpoints, such as the BRMA model (\ref{brma-w})--(\ref{brma-b}), in a special case of data structure when there are only two treatments in the network, as detailed in Appendix \ref{supl.models}.

\newpage
\begin{table}[t]
%\footnotesize
%\small
\centering
\begin{tabular}{lcccccc}
& \multicolumn{6}{c}{Assumptions for the between-studies variance-covariance matrix}  \\ \cline{2-7}
            &  varying  &  second   &             & homogeneity  & prior &  \\
            &  variance-  & order & exchangeable &  of variance-  & distributions & exchangeable \\
            & -covariance  & consistency & ancillary & -covariance   & on  basic  & treatments \\
model      &  matrix     &  Eqs & parameters         &  matrix &  parameters  &   Eqs \\
&  Eq. (\ref{eq-bvnma-b-het}) &  (\ref{eq-sec-ord-consist})--(\ref{eq-sec-ord-consist2}) & Eq. (\ref{eq-var-sim}) & Eq. (\ref{eq-bvnma-b-hom}) &  Eq. (\ref{eq-prior-d}) & (\ref{eq-exchange-1})--(\ref{eq-exchange-2})  \\
\hline
1a    & \checkmark    &             &   &   &   \checkmark &  \\
1b    & \checkmark    & \checkmark  &  &   &   \checkmark &    \\
1c    & \checkmark    & \checkmark  & \checkmark &   &   \checkmark &    \\
1d    &               &      NA       &   & \checkmark  &   \checkmark &  \\ \hline
2a    & \checkmark    &             &  &   &    &   \checkmark  \\
2b    & \checkmark    & \checkmark  &  &   &    &   \checkmark  \\
2c    & \checkmark    & \checkmark  & \checkmark  &   &    &   \checkmark  \\
2d    &               &     NA       &  &   \checkmark &    &   \checkmark   \\
\hline
\end{tabular}
\caption{Components of each model in terms of the assumptions made about the between-studies variance-covariance (v-c) matrix.}
\label{tab-models}
\end{table}

\subsubsection{Strategies for model comparison}
\label{sec-best}
To make comparisons between the models, in the first instance the between-studies correlations (as well as the mean effects and the heterogeneity parameters) are obtained and compared across the models.
 Following this, predicted values (and corresponding credible intervals (CrIs)) of the treatment effects on the final outcome are compared to the observed estimates (and corresponding confidence intervals (CIs)) in take-one-out cross-validation procedure.
In one study at a time, the estimate of the treatment effect on the final outcome $Y_{2i}$ is removed (and treated as missing at random) and then this treatment effect is predicted from the treatment effect on the surrogate endpoint, conditional of the data on both outcomes from all the remaining studies in the meta-analysis.
The standard deviation of the predicted effect $\hat{Y}_{2i}$ is equal to
$\sqrt{\sigma_{2i}^2+var(\hat{\mu}_{2(kl)i}\vert Y_{1i},\sigma_{1i},Y_{1(-i)},Y_{2(-i)})}$, where $Y_{1(-i)}$ and $Y_{2(-i)}$ denote the data from the remaining studies without the validation study $i$ (the treatment contrast subscripts, present only in the network meta-analysis, have been dropped here).

The models were compared with respect to the predicted effects by investigating a number of statistics obtained from each model: a) proportion of the CI of the observed estimate of the effect on the final outcome overlapping with the CrI of the predicted effect, $p_{overlap}$, b) mean absolute difference between observed estimate of mean effect and predicted mean effect on the final outcome across studies, $\vert m_{obs}-m_{pred} \vert$, c) ratio of the width of intervals; the width of CrI of the predicted effect to the width of the CI of the observed effect (the estimate),  $w_{pred}/w_{obs}$, d) percentage reduction in uncertainty measured by the width of the CrI of the predicted effect when using a mvNMA model, $w_{NMA}$, compared to the width of the CrI obtained from BRMA $w_{BRMA}$; $\%red=100\frac{w_{BRMA}-w_{NMA}}{w_{BRMA}}$, and e) a new statistic measuring overall performance of a model giving a higher score for models resulting in large $p_{overlap}$ with penalty for overly inflated predicted interval:
$\pi=\frac{p_{overlap}}{w_{pred}/w_{obs}}$, which is always positive and less than one.

\newpage
\section{Results for aCRC data}
\label{sec.acrc}
\begin{figure}[p]
\centering
\includegraphics[scale=0.15]{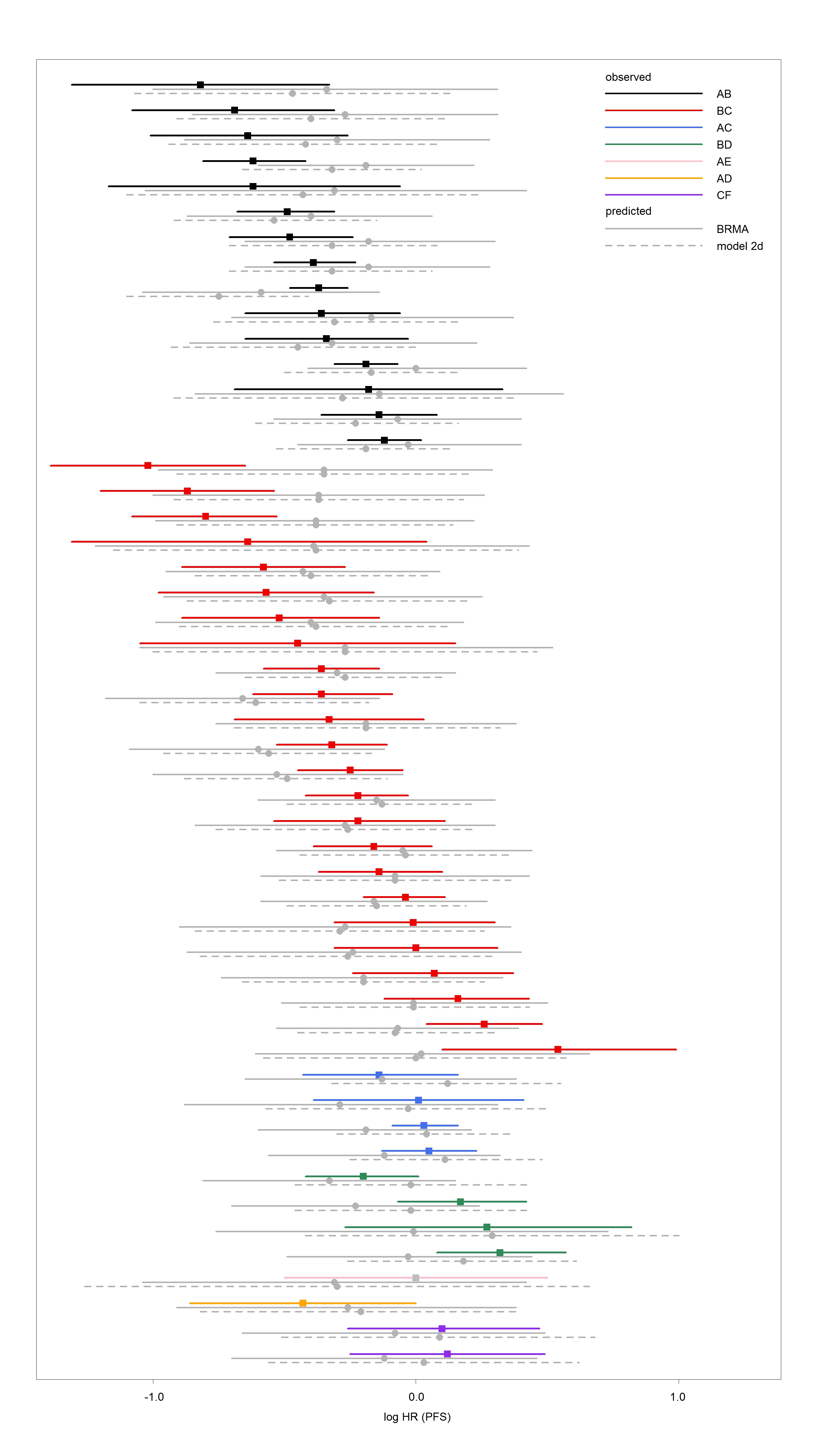}
\caption{Predicted effects obtained from BRMA and model 2d along with the observed estimates of the effects on PFS for aCRC data}
\label{acrc2.pred}
\end{figure}

In this section we present results of applying all models (BRMA, 1a-d and 2a-d) to the illustrative example in aCRC, introduced in Section \ref{sec.ill}.
Table \ref{varcov.acrc} shows the between-studies correlations obtained from all the models  applied to the aCRC data.
The correlations are the parameters of the primary interest as they tell us about the strength of the surrogate relationships (see Sections \ref{sec.sur.cri1} and \ref{sec.sur.cri2}).
Additional results, the average effects and heterogeneity parameters, are included in Appendix \ref{supl.results.acrc}.
When applying BRMA to data from all studies (regardless of the treatment contrast), the between-studies correlation is negative,
-0.73 (95\% CrI: -0.89, -0.49), indicating that treatment increasing the odds of tumour response leads to reduced progression rate.
As shown in the scatter plot in Figure \ref{fig_acrc_scatter}, the distributions of the treatment effects on the two outcomes: log OR of TR (surrogate endpoint) and log HR of PFS (final outcome), representing their association patterns, vary across the treatment contrasts.
When applying NMA models 1a-c and 2a-c to the data, the between-studies correlations differ across treatment contrasts, with the highest correlation obtained for treatment contrast AC (the results are presented only for those contrasts for which at least four studies were available, the remaining results  are included in Appendix \ref{supl.results.acrc}).
\begin{sidewaystable}[p]
\centering
\begin{tabular}{lcccccc}
 & \multicolumn{4}{c}{\emph{within-treatment surrogate relationship}} &&\\
    \cline{2-5}
model & AB & AC & BC & BD & $\rho_t$ & DIC\\ \hline
\hline
BRMA&\multicolumn{4}{c}{-0.73 (-0.89, -0.49)}& NA & -10.6\\
1a&-0.52 (-0.92, 0.07)&-0.83 (-0.97, -0.54)&0.00 (-0.90, 0.88)&-0.27 (-0.94, 0.72)& NA & -12.8\\
1b&-0.69 (-0.96, -0.18)&-0.83 (-0.97, -0.57)&-0.29 (-0.91, 0.67)&-0.28 (-0.89, 0.58)& NA & -12.3\\
1c&-0.71 (-0.96, -0.25)&-0.82 (-0.97, -0.53)&-0.24 (-0.89, 0.69)&-0.3 (-0.89, 0.53) & NA & 4.95\\
1d&\multicolumn{4}{c}{-0.81 (-0.96, -0.59)}& NA & -14.0\\
2a&-0.53 (-0.91, 0.05)&-0.82 (-0.97, -0.52)&-0.04 (-0.9, 0.87)&-0.31 (-0.94, 0.61)  & -0.35 (-0.93, 0.53) & -14.6\\
2b&-0.70 (-0.95, -0.19)&-0.83 (-0.97, -0.56)&-0.28 (-0.91, 0.68)&-0.29 (-0.9, 0.56) & -0.34 (-0.92, 0.56) & -14.1\\
2c&-0.71 (-0.96, -0.27)&-0.81 (-0.97, -0.52)&-0.25 (-0.89, 0.68)&-0.3 (-0.89, 0.54) & -0.34 (-0.91,0.56) & 3.05\\
2d&\multicolumn{4}{c}{-0.82 (-0.97, -0.58)} & -0.37 (-0.92,0.54) & -15.3\\
\hline
\end{tabular}
\caption{Between-studies correlations corresponding to the within-treatment surrogate relationship for each model, $\rho_t$ -- across-treatment correlations obtained from the models allowing for exchangeability, and DIC values corresponding to each model fitted to aCRC data. Where only one value is given for the between-studies correlation within a treatment contrast (models BRMA, 1d and 2d), the parameters are common across the treatment contrasts.
A -- chemotherapy alone, B -- anti-VEGF therapies + chemotherapy,
C -- EGFRi  + chemotherapy, D -- EGFRi + anti-VEGF therapies + chemotherapy}
\label{varcov.acrc}
\end{sidewaystable}
When applying model 1b, assuming  second order consistency and hence imposing additional constraints on the between-studies variance-covariance matrices, the correlations were estimated with higher precision compared to those obtained from model 1a. Applying model 1c slightly inflated the credible intervals of the correlations for some of the treatment contrasts (AC and BC) and reduced them for others (AB and BD) (compared to model 1b), which is likely due to the assumption of similar variances for the  effects in each treatment arm not being reasonable for these data.
The DIC value, shown in the right-hand-side column of the table, is also relatively high for  model 1c and indicates poor model fit.
A similar pattern is observed across models 2a-c.
When assuming homogeneous variance-covariance matrices in models 1d and 2d, the between-studies correlations are high and obtained with higher precision compared to the correlation obtained from BRMA.
This is likely due to the borrowing of strength across treatment contrasts, by adding information from the indirect comparisons.
Based on the DIC criteria, models 1d, 2a, 2b and 2d appear to fit the data best, although models 1a and 1b appear to have a comparable fit.
Prediction of the treatment effect on PFS (the final outcome) from the treatment effect on TR (the surrogate endpoint) for the treatment contrast AE,  for which  only one study was available (new treatment scenario), was not possible for models 1a--d, due to lack of data to estimate the surrogate relationship.
Models 2a--d, assuming exchangeability of the average effects in each treatment arm  across treatment contrasts, allowed us to make such a prediction.
A forest plot of the observed and predicted (from BRMA and model 2d) effects on PFS is included in Figure \ref{acrc2.pred}.
The advantage of using bvNMA methods varied across the treatment contrasts, depending on the surrogacy relationships for each contrast.
Table \ref{varcov.acrc} also shows the across-treatment correlation $\rho_t$ indicating, consistently across models 2a--d, a weak surrogate relationship across the treatment contrasts.

Table \ref{acrc-sbest} shows statistics for model comparison, introduced in Section \ref{sec-best}, for each treatment contrast where at least four studies were available and across all the studies. For example, for the treatment contrast AB, the use of bvNMA improved the precision of the predictions in terms of the point estimate, reducing the bias from 0.23 (obtained from BRMA) to between 0.16 and 0.18.
All bvNMA methods gave reduced predicted intervals for the treatment effect on PFS (from the effect on TR) compared to those obtained from BRMA, by between 11.2\% and 16.5\%. For contrasts AC and BC, model 1a did not contribute to reduced uncertainty of predictions, but additional assumptions of second order consistency (models 1b and 2b) and additional borrowing of strength (models 1c and 2c) or assumption of homogeneity of the correlations and heterogeneity parameters (models 1d and 2d) led to improved precision.
Results from a sensitivity analysis of the data with outlying observations removed are presented in Appendix \ref{sec.sens.anal}.
For the treatment contrast BD, both sets of models 1a--b and 2a--b resulted in  inflated predicted intervals. The final part of Table \ref{acrc-sbest} lists the average statistics across all treatments.

\begin{table}[p]
\centering
\begin{tabular}{lccccc}
&   $p_{overlap}$  &  $\vert m_{obs}-m_{pred} \vert$  &  $w_{pred}/w_{obs}$ & $\pi$ & $\% red.$ \\ \hline
\hline
\multicolumn{6}{l}{AB}\\
BRMA&0.89&0.23&2.2&0.45&\\
model 1a&0.91&0.18&1.88&0.54&14.64\\
model 1b&0.91&0.17&1.95&0.52&11.22\\
model 1c&0.89&0.17&1.82&0.54&16.54\\
model 1d&0.89&0.16&1.85&0.53&14.18\\
model 2a&0.91&0.18&1.91&0.53&13.55\\
model 2b&0.9&0.17&1.95&0.52&11.25\\
model 2c&0.89&0.17&1.83&0.54&16.33\\
model 2d&0.89&0.16&1.84&0.53&14.76\\
\hline
\multicolumn{6}{l}{AC}\\
BRMA&0.92&0.24&1.91&0.5&\\
model 1a&0.95&0.23&1.93&0.51&-1.27\\
model 1b&0.92&0.24&1.8&0.52&5.42\\
model 1c&0.88&0.24&1.7&0.53&10.41\\
model 1d&0.86&0.24&1.64&0.54&13.28\\
model 2a&0.94&0.23&1.93&0.51&-1.02\\
model 2b&0.91&0.24&1.79&0.52&6.05\\
model 2c&0.87&0.24&1.69&0.53&10.88\\
model 2d&0.86&0.24&1.63&0.54&14.06\\
\hline
\multicolumn{6}{l}{BC}\\
BRMA&0.97&0.17&2.24&0.47&\\
model 1a&1.00&0.08&3.05&0.4&-30.83\\
model 1b&0.97&0.09&2.12&0.5&6.22\\
model 1c&0.95&0.09&1.83&0.56&17.91\\
model 1d&0.95&0.1&1.9&0.53&14.4\\
model 2a&1.00&0.08&2.93&0.41&-26.52\\
model 2b&0.97&0.08&2.09&0.51&7.49\\
model 2c&0.95&0.07&1.79&0.57&19.81\\
model 2d&0.95&0.09&1.89&0.54&14.87\\
\hline
\multicolumn{6}{l}{BD}\\
BRMA&0.83&0.29&1.86&0.46&\\
model 1a&1.00&0.23&4.58&0.24&-144.64\\
model 1b&1.00&0.23&3.92&0.27&-109.09\\
model 1c&0.84&0.25&1.8&0.49&2.7\\
model 1d&1.00&0.14&1.79&0.58&3.52\\
model 2a&1.00&0.21&3.67&0.29&-97.09\\
model 2b&1.00&0.21&3.11&0.34&-65.92\\
model 2c&0.84&0.24&1.67&0.52&9.54\\
model 2d&1.00&0.13&1.73&0.59&6.57\\
\hline
\multicolumn{6}{l}{All}\\
BRMA&0.91&0.23&1.99&0.49&0\\
model 1a&0.91&0.2&2.38&NA&-22.58\\
model 1b&0.89&0.2&2.21&NA&-14.84\\
model 1c&0.86&0.2&1.7&NA&13.97\\
model 1d&0.86&0.19&1.68&NA&14.86\\
model 2a&0.95&0.2&2.36&0.46&-22.76\\
model 2b&0.93&0.2&2.14&0.49&-11.77\\
model 2c&0.89&0.2&1.73&0.54&11.52\\
model 2d&0.9&0.19&1.72&0.55&12.06\\
\hline
\end{tabular}
\caption{\label{acrc-sbest}Comparison of models based on aCRC data by treatment contrast.}
\end{table}

\newpage
\section{Illustration using simulated data}
\label{sec.ill.sim}
\subsection{Data simulation}
\begin{figure}[h]
\centering
\includegraphics[scale=0.25]{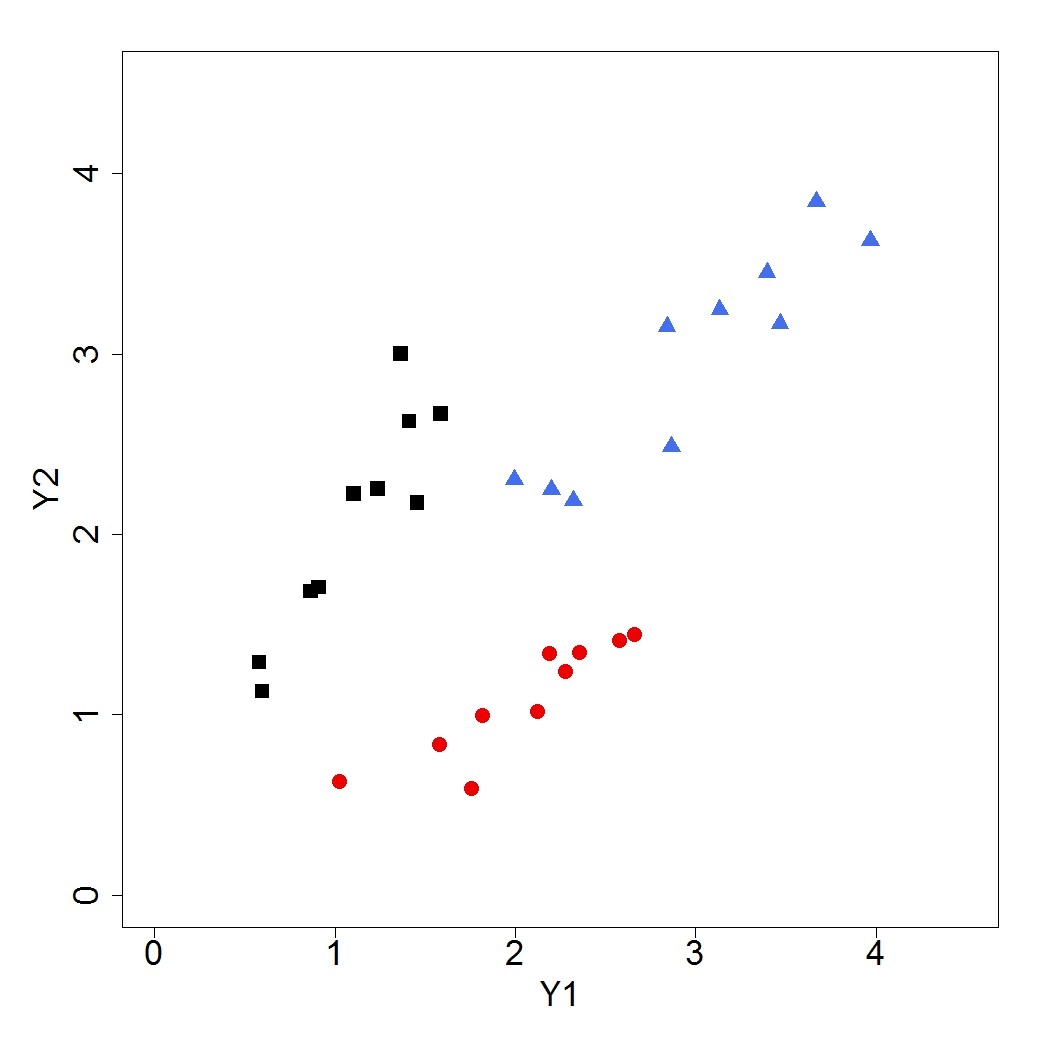}\hspace{0.25cm}
\includegraphics[scale=0.25]{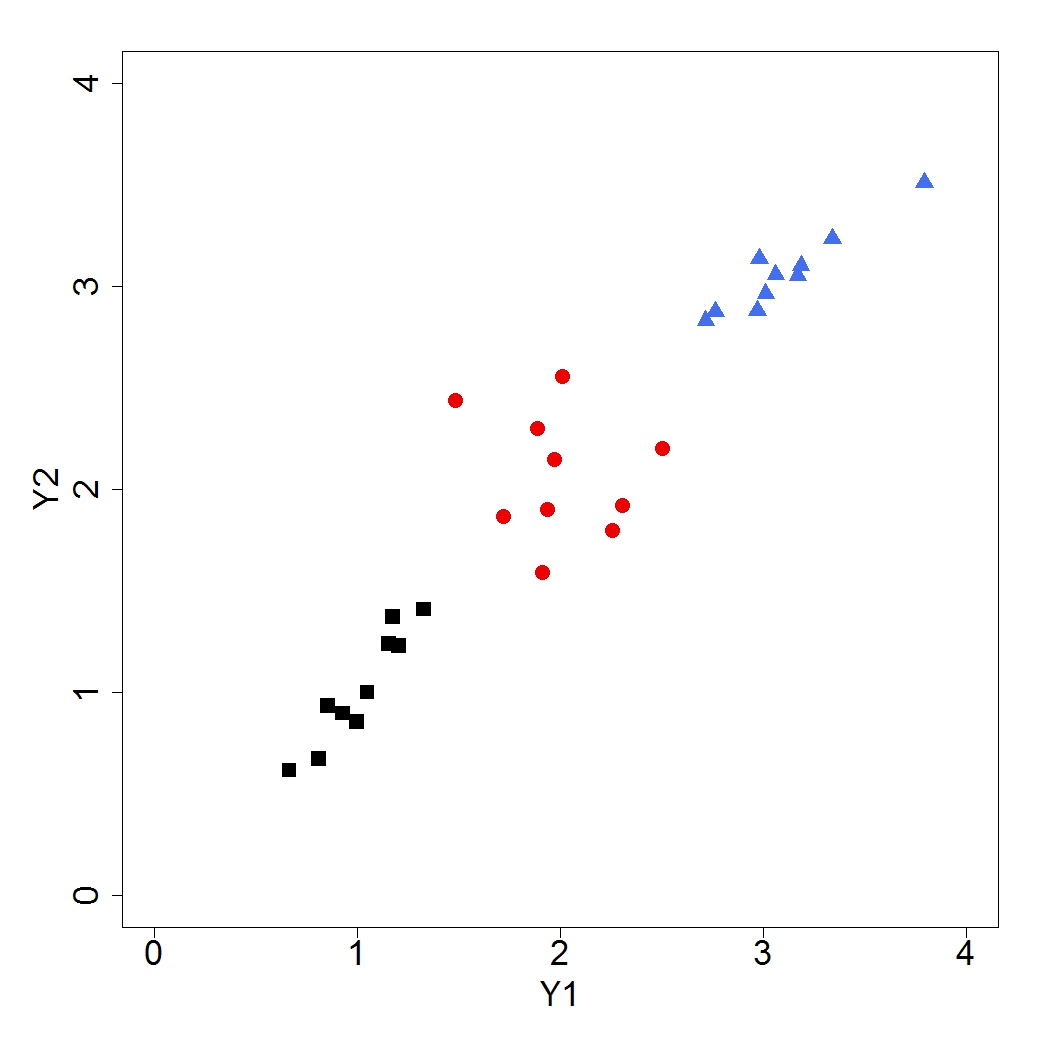}\hspace{0.25cm}\\
\begin{tikzpicture}[nodes={draw, circle}, line width=1.2pt]
\centering
[inner sep=3mm]
\path node at ( 0,1.6) [shape=circle] (f) {A}
node at ( -1,0) [shape=circle] (s1) {B}
node at ( 1,0) [shape=circle] (s2) {C}
node at (0,-1) [shape=circle,draw=white] (s4) {} %ghost node
(s1) edge[black]  (f)
(s2) edge[blue!80!green] (f)
(s1) edge[red] (s2);
\end{tikzpicture}
\vspace{-0.5cm}
\caption{Scatter plots of the artificial data simulated under scenario 1 (top left) and scenario 2 (top right) and network diagram corresponding to the structure of data for both scenarios.}
\label{scatters}
\end{figure}
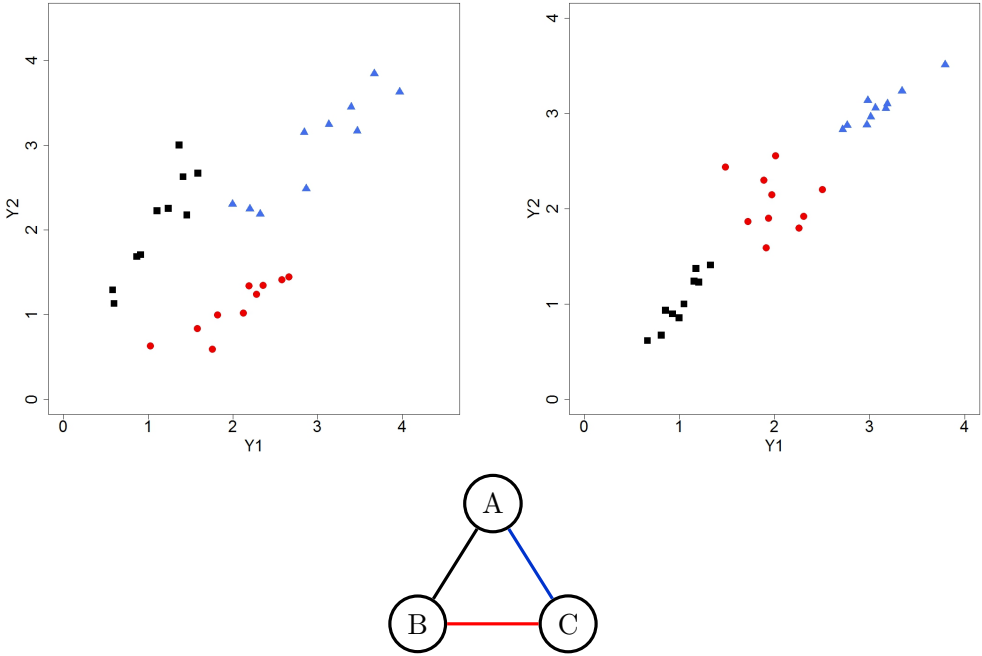

To demonstrate scenarios where use of bvNMA methods has an advantage over the standard surrogacy models, data were simulated under different assumptions.
%In particular, we simulated data where the overall association pattern across all studies and treatments was weak but the surrogate relationships were strong within treatment contrasts or where the surrogacy relationship was strong only for some treatment contrasts, detectable by  mvNMA but not by BRMA.
In particular, we simulated data where the surrogate pattern  across all studies and treatments differed from the patterns within treatment contrasts, which is detectable by  mvNMA but not by BRMA.
The treatment effects on two outcomes  were simulated from a bivariate normal distribution:
\begin{equation}
\left(
\begin{array}{c}
Y_{1kli}\\
Y_{2kli}\\
\end{array}
\right) \sim \rm{N}
\left\{
\left(
\begin{array}{c}
d_{1kl}\\
d_{2kl}\\
\end{array}
\right),
%\mathbf{\Sigma_i}
%\right), \;
%\mathbf{\Sigma_i}=
\left(
\begin{array}{cc}
\sigma_{1kli}^2 + \tau_{1kl}^2 & \sigma_{1kli}\sigma_{2kli}\rho_{wkli} + \tau_{1kl}\tau_{2kl}\rho_{1kl,2kl}\\
\sigma_{1kli}\sigma_{2kli}\rho_{wkli} + \tau_{1kl}\tau_{2kl}\rho_{1kl,2kl} & \sigma_{2kli}^2 + \tau_{2kl}^2
\end{array}
\right)
\right\},
\nonumber
\end{equation}
as in  model 1a.
Two sets of network data were generated, each consisting of 30 studies, three treatments and three treatment contrasts with 10 studies per contrast (AB, BC and AC), under different scenarios (illustrated in Figure \ref{scatters}).

\textbf{Scenario 1} was simulated assuming weak surrogacy when ignoring treatment contrasts but strong surrogacy within each treatment contrast, with the following parameters: $\mathbf{d}_{AB}=(1,2)$, $\mathbf{d}_{BC}=(2,1)$, $\mathbf{d}_{AC}=(3,3)$;
    $\sigma_{jAB(AC,BC)i}\sim Unif(0.15, 0.25)$, j=1,2;  $\rho_{wABi}=\rho_{wBCi}=\rho_{wACi}=0.6$;
     $\tau_{1AB}=0.3$, $\tau_{1BC}=\tau_{1AC}=0.6$, $\tau_{2BC}=0.3$, $\tau_{2AB}=\tau_{2AC}=0.6$; $\rho_{AB}=\rho_{AC}=\rho_{BC}=0.98$.

\textbf{Scenario 2} was simulated assuming a strong surrogacy relationship when ignoring treatment contrasts and a mixture of either weak or strong relationships within each treatment contrast, with the following parameters:   $\mathbf{d}_{AB}=(1,1)$, $\mathbf{d}_{BC}=(2,2)$, $\mathbf{d}_{AC}=(3,3)$; $\sigma_{jAB(AC,BC)i} \sim Unif(0.05, 0.15)$, j=1,2; $\rho_{wABi}=\rho_{wBCi}=\rho_{wACi}=0.98$;  $\tau_{1AB}=0.2$, $\tau_{1BC}=0.25$, $\tau_{1AC}=0.3$, $\tau_{2AB}=0.3$, $\tau_{2BC}=0.25$, $\tau_{2AC}=0.2$; $\rho_{AB}=\rho_{AC}=0.98$ and $\rho_{BC}=0$ (strong surrogacy relationships for treatment contrasts AB and AC but no relationship for BC).
%\newpage

\subsection{Results of the analysis of the simulated data}
\subsubsection{Scenario 1}
\begin{figure}[p]
\centering
a) \includegraphics[scale=0.175]{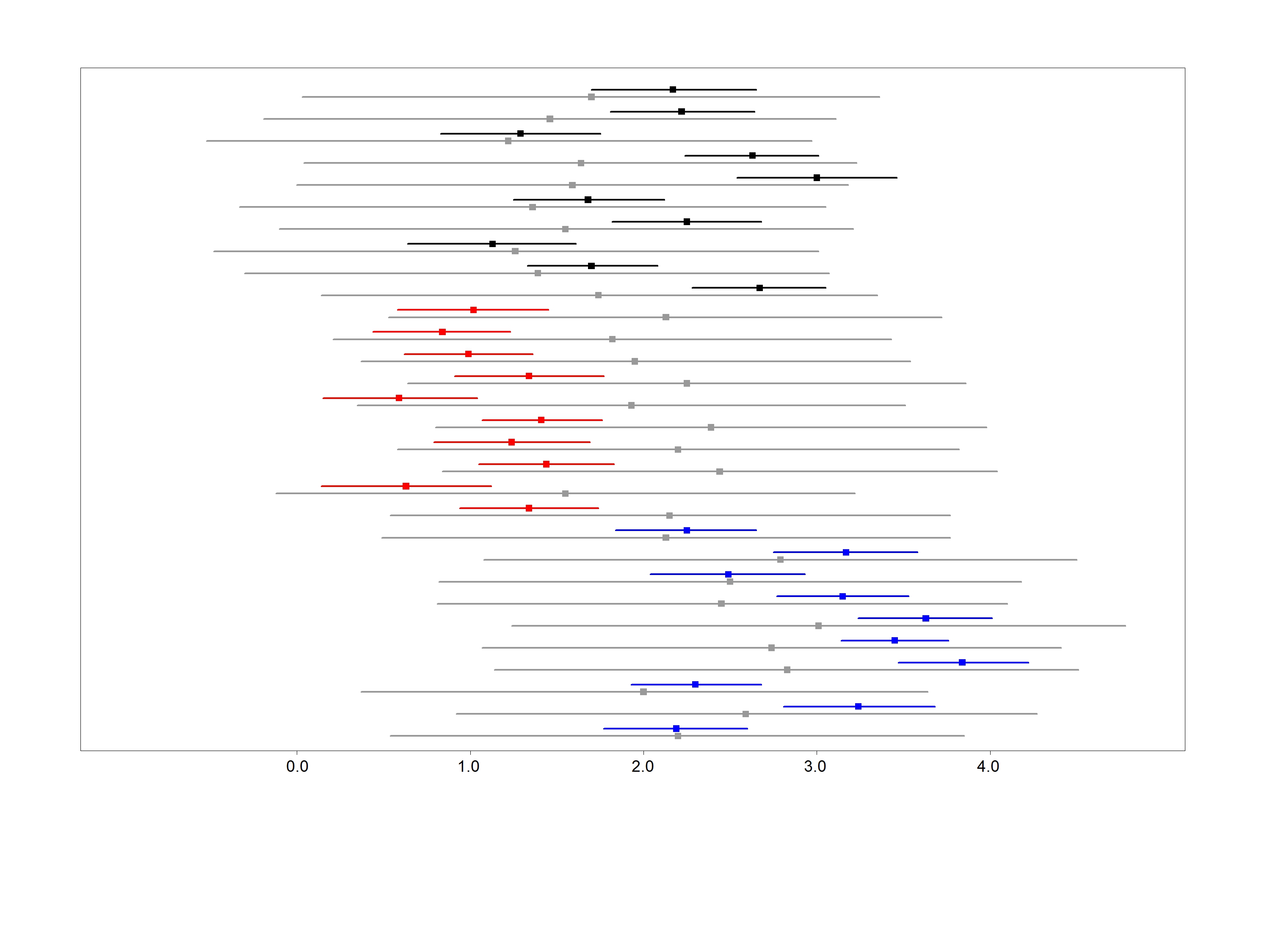}\\
b) \includegraphics[scale=0.175]{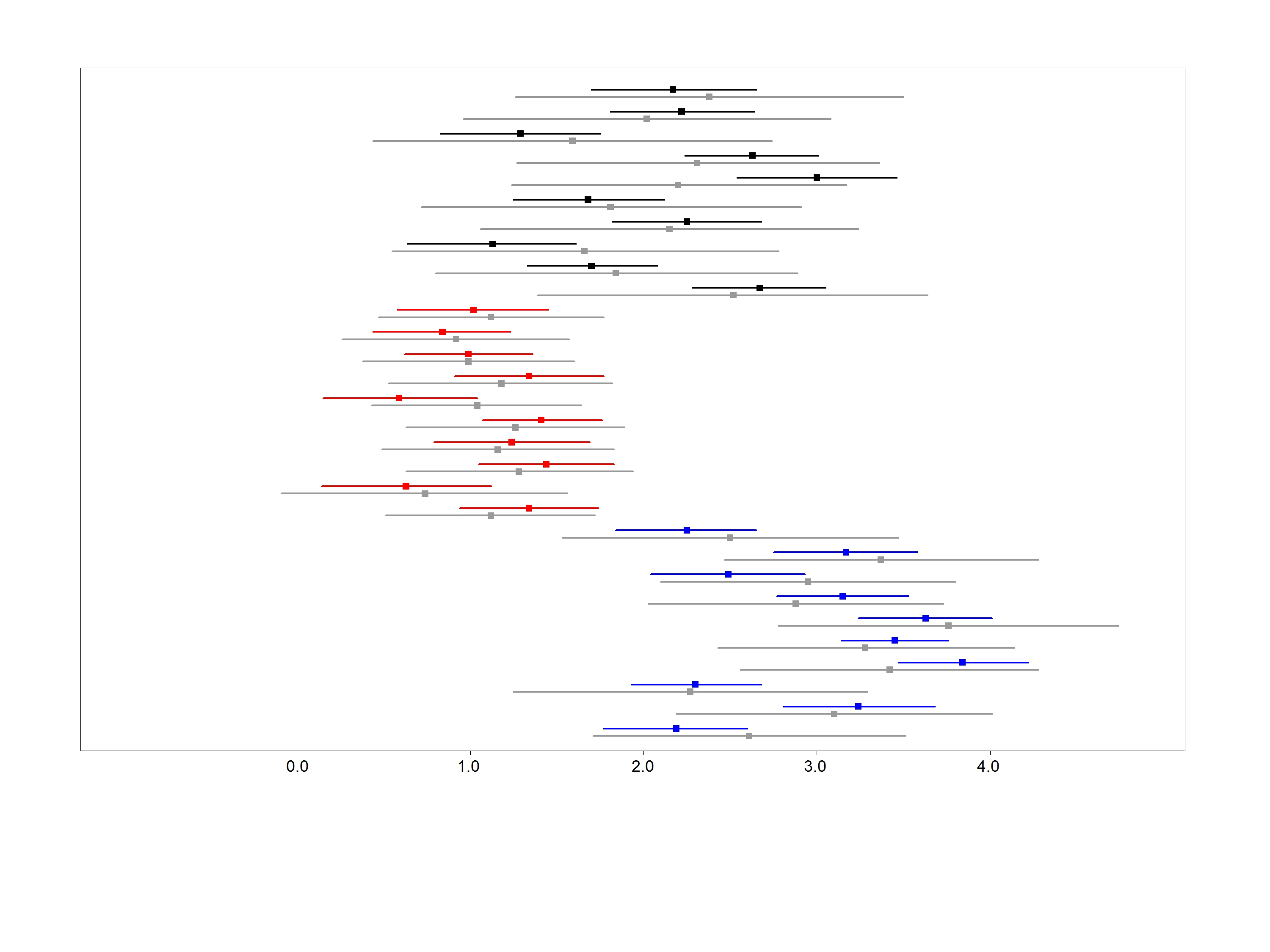}
\caption{Predicted effect (gray) obtained from the cross validation procedure are presented along the observed effects: black (B vs A), red (C vs B) and blue (C vs A), obtained from (a) BRMA  and (b) model 1a for  data simulated under scenario 1.}
\label{sc1.pred}
\end{figure}

Table \ref{varcov.sim1} shows the between-studies correlations obtained by applying all models to the data simulated under scenario 1.
The between-studies correlation obtained from BRMA (across all studies) was not very high: 0.57 (95\% CrI: 0.27, 0.79).
Bivariate NMA with the covariance matrix varying across treatment contrasts models the data in more detail and reveals strong correlation between outcomes within the treatment contrasts, namely 0.79 (0.26, 0.99) for treatment contrast AB, 0.74 (0.09, 0.99) for BC and 0.85 (0.46, 0.99) for AC when using model 1a.
Placing additional constraints on the covariance matrix by assuming  second order consistency in models 1b and 1c reduced uncertainty around the correlations.
Model 1d resulted in a common correlation obtained with the highest precision, however it did not take into account the differences in the between-studies variances across the treatment contrasts, in contrast to models 1a-c as shown in Appendix \ref{supl.sim1}.
The right-hand-side  column of Table \ref{varcov.sim1} shows the across-treatment correlations $\rho_t$ indicating a weak across-treatment surrogate relationship. This is due to the differences in the surrogacy patterns across the treatment contrasts as well as the small number of treatment contrasts.
Table \ref{sbest.sc1} shows a range of statistics (described in Section \ref{sec-best}) comparing the models in terms of their value in predicting the treatment effect on the final outcome from the treatment effect measured on the surrogate endpoint in a cross-validation procedure.
The large width of the predicted interval obtained from BRMA compared to the width of the CI of the observed treatment effect estimate is due to high uncertainty, but the ratio $w_{pred}/w_{obs}$ is reduced when using NMA models. Predicted intervals obtained from NMA models are between 47\% and 57\% narrower compared to those obtained from BRMA. The distance between the point estimate of the observed   effect from the predicted effect is also much reduced when using NMA models compared to BRMA.
Figure \ref{sc1.pred} shows predicted effects obtained (a)  using BRMA and (b) from model 1a obtained with higher precision.
\begin{table}[h]
\centering
\begin{tabular}{lcccc}
&\multicolumn{3}{c}{\emph{within-treatment surrogate relationship}} &\\ \cline{2-4}
model & AB & BC & AC & $\rho_t$ \\ \hline
BRMA& \multicolumn{3}{c}{0.57 (0.27, 0.79)} & NA \\
model 1a&0.79 (0.26, 0.99)&0.74 (0.09, 0.99)&0.85 (0.46, 0.99)& NA\\
model 1b&0.88 (0.55, 0.99)&0.74 (0.22, 0.97)&0.9 (0.66, 0.99)& NA\\
model 1c&0.87 (0.54, 0.99)&0.71 (0.17, 0.96)&0.89 (0.64, 0.99)& NA\\
model 1d&\multicolumn{3}{c}{0.92 (0.77, 0.99)}& NA\\
model 2a&0.78 (0.19, 0.99)&0.74 (0.12, 0.98)&0.85 (0.48, 0.99) & 0.46 (-0.31, 0.93) \\
model 2b&0.88 (0.56, 0.99)&0.74 (0.23, 0.97)&0.9 (0.65, 0.99) & 0.47 (-0.30, 0.93) \\
model 2c&0.86 (0.54, 0.99)&0.72 (0.18, 0.96)&0.89 (0.64, 0.99) & 0.48 (-0.28, 0.94)\\
model 2d&\multicolumn{3}{c}{0.92 (0.76, 0.99)}  & 0.44 (-0.33, 0.93)\\ \hline
\end{tabular}
\caption{\label{varcov.sim1}Between-studies correlations for each model under simulation scenario 1. Where only one value is given (models BRMA, 1d and 2d), the parameters are common across the treatment contrasts. $\rho_t$ is the  correlation corresponding to the across-treatment surrogate relationship.}
\end{table}
\begin{table}[h]
\centering
\begin{tabular}{lccccc}
&   $p_{overlap}$  &  $\vert m_{obs}-m_{pred} \vert$  &  $w_{pred}/w_{obs}$ & $\pi$ & $\% red.$ \\ \hline
\hline
BRMA&0.98&0.69&4.03&0.25&\\
model 1a&0.98&0.23&2.15&0.48&46.74\\
model 1b&0.97&0.22&1.91&0.52&52.53\\
model 1c&0.95&0.24&1.77&0.55&55.92\\
model 1d&0.94&0.24&1.69&0.56&57.88\\
model 2a&0.98&0.23&2.15&0.48&46.67\\
model 2b&0.97&0.22&1.91&0.52&52.57\\
model 2c&0.96&0.24&1.78&0.55&55.76\\
model 2d&0.94&0.24&1.69&0.56&57.97\\
\hline
\end{tabular}
\caption{\label{sbest.sc1}Comparison of models based on simulation scenario 1.}
\end{table}
\subsubsection{Scenario 2}
\begin{figure}[p]
\centering
a) \includegraphics[scale=0.175]{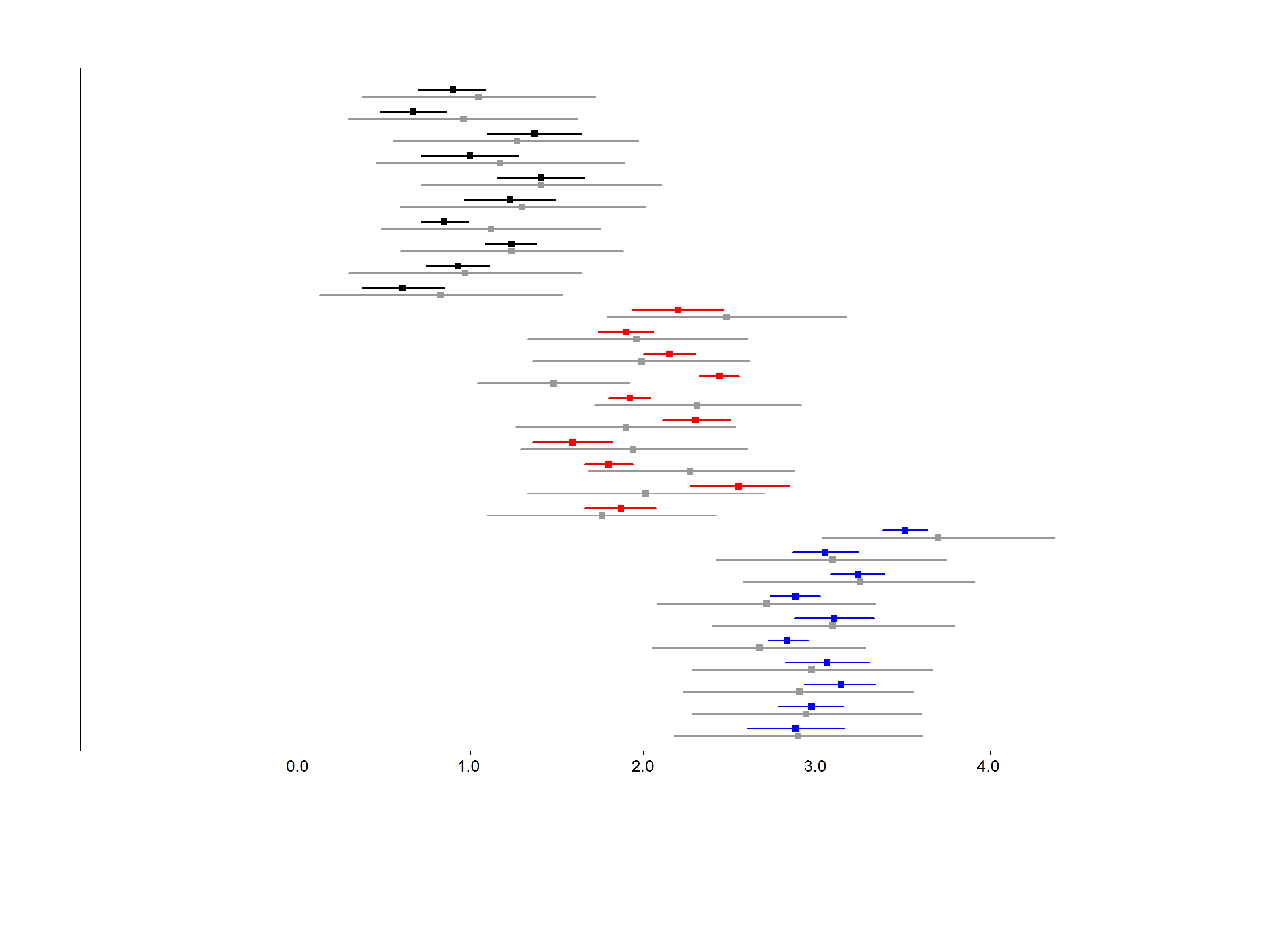}\\
b) \includegraphics[scale=0.175]{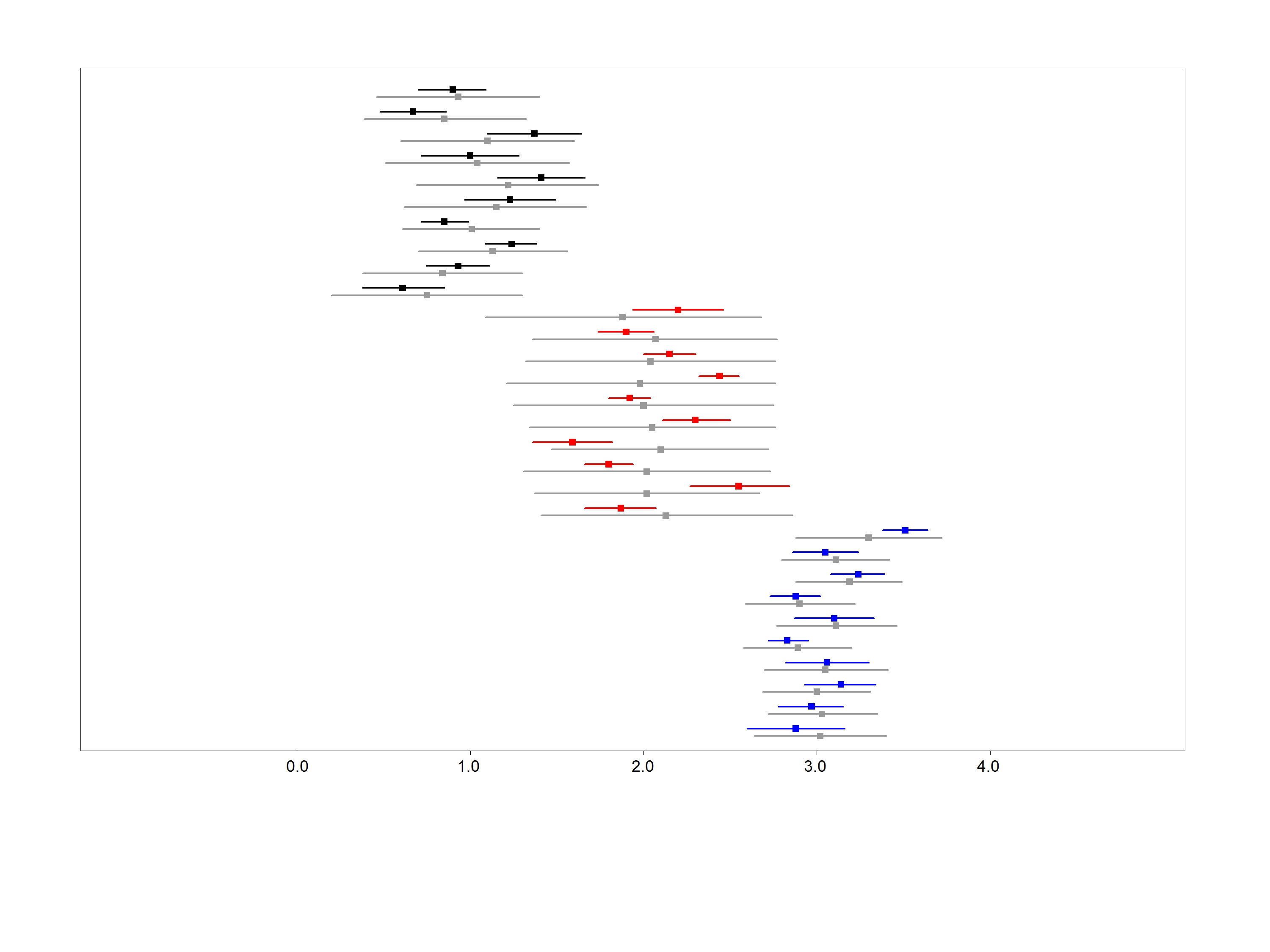}
\caption{Predicted effect (gray) obtained from the cross validation procedure are presented along the observed effects: black (B vs A), red (C vs B) and blue (C vs A), obtained from (a) BRMA and (b) model 1a for  data simulated under scenario 2.}
\label{sc2.pred}
\end{figure}

Table \ref{varcov.sim2} shows the between-studies correlations for the data simulated under scenario 2.
The overall correlation obtained from BRMA is high: 0.94 (95\% CrI: 0.88, 0.98).
Bivariate NMA with the variance-covariance matrix varying across treatment contrasts models the data in more detail and reveals no correlation between the treatment effects on the two outcomes within the BC treatment contrast.
Figure \ref{sc2.pred} shows the predicted effects obtained from the cross-validation using (a) BRMA and (b) NMA model 1a. When using the NMA model, predictions are obtained with higher precision  for contrasts AB and AC, but not BC where there was no association between the effects on the two outcomes and which is reflected by the wide predicted intervals.
The across-treatment correlations $\rho_t$ in the right-hand-side  column of Table \ref{varcov.sim2} are obtained with high uncertainty due to the small number of treatment contrasts to estimate the correlation.
Table \ref{sbest.sc2} shows the statistics for the model comparison in terms of their predictive value, obtained from the cross-validation procedure.
Similarly as in scenario 1, the large ratio, $w_{pred}/w_{obs}$,  comparing the width of the predicted interval obtained from BRMA with the width of the CI of the observed treatment effect estimate is reduced when using the NMA models. Predicted intervals obtained from NMA models are between 19.6\% and 29.3\% narrower compared to those obtained from BRMA.
The distance between the point estimate of the observed   effect and the predicted effect is slightly reduced when using  NMA models 1a-c and 2a-c compared to BRMA. These improvements, on average, are not as strong as in scenario 1, due to poor association for the treatment contrast BC. When investigating these statistics within the treatment contrasts, the improvement is largest for contrast AC where the correlation was highest (these results are presented in Appendix \ref{supl.sim2}).
\begin{table}[h]
\centering
\begin{tabular}{lcccc}
&\multicolumn{3}{c}{\emph{within-treatment surrogate relationship}} &\\ \cline{2-4}
model & AB & BC & AC & $\rho_t$ \\
\hline
BRMA&\multicolumn{3}{c}{0.94 (0.88, 0.98)} & NA\\
model 1a&0.81 (0.34, 0.99)&-0.19 (-0.73, 0.43)&0.87 (0.49, 0.99) & NA\\
model 1b&0.78 (0.2, 0.99)&-0.05 (-0.6, 0.53)&0.8 (0.26, 0.99) & NA\\
model 1c&0.77 (0.22, 0.98)&-0.11 (-0.62, 0.43)&0.79 (0.3, 0.99) & NA\\
model 1d&\multicolumn{3}{c}{0.39 (0, 0.69)} & NA\\
model 2a&0.81 (0.32, 0.99)&-0.2 (-0.75, 0.44)&0.87 (0.47, 0.99) & 0.56 (-0.25, 0.99) \\
model 2b&0.77 (0.19, 0.99)&-0.04 (-0.59, 0.54)&0.79 (0.26, 0.99) & 0.58 (-0.22, 0.99)\\
model 2c&0.78 (0.25, 0.99)&-0.11 (-0.62, 0.42)&0.79 (0.29, 0.98) & 0.59 (-0.26, 0.99)\\
model 2d&\multicolumn{3}{c}{0.39 (0, 0.69)} & 0.57 (-0.27, 0.99)\\
\hline
\end{tabular}
\caption{\label{varcov.sim2}Between-studies correlations for each model under simulation scenario 2. Where only one value is given (models BRMA, 1d and 2d), the parameters are common across the treatment contrasts. $\rho_t$ is the  correlation corresponding to the across-treatment surrogate relationship.}
\end{table}

%\newpage

\begin{table}[h]
\centering
\begin{tabular}{lccccc}
&   $p_{overlap}$  &  $\vert m_{obs}-m_{pred} \vert$  &  $w_{pred}/w_{obs}$ & $\pi$ & $\% red.$ \\ \hline
\hline
BRMA&0.9&0.2&3.55&0.27&\\
model 1a&0.97&0.16&2.84&0.41&20.2\\
model 1b&0.95&0.17&2.76&0.38&21.86\\
model 1c&0.93&0.17&2.5&0.41&29.16\\
model 1d&0.9&0.2&2.87&0.33&19.5\\
model 2a&0.98&0.16&2.83&0.41&20.32\\
model 2b&0.95&0.17&2.77&0.38&21.86\\
model 2c&0.93&0.17&2.49&0.41&29.34\\
model 2d&0.9&0.2&2.87&0.33&19.61\\
\hline
\end{tabular}
\caption{\label{sbest.sc2}Comparison of models based on simulation scenario 2.}
\end{table}

\newpage
\section{Discussion}
We have developed bivariate network meta-analytic models for surrogate endpoint evaluation to allow for a more detailed modelling of surrogate relationships within and across treatment contrasts.
% or treatment classes.
This methodology can help to disentangle information about surrogate relationships in data scenarios where such relationships vary across treatment contrasts and are distinct in comparison with an association pattern across treatments.
Two types of surrogacy have been described by the models: surrogacy across patient populations for a given treatment contrast and surrogacy across treatments.
The models will allow analysts to make predictions of the treatment effect on the final clinical outcome from the observed effect on a surrogate endpoint in a new study investigating the effectiveness of an existing treatment in a new setting or a new treatment.

There are some limitations to the models presented here. For simplicity, we focused on the  models for data from two-arm studies. The methods can be extended to model multi-arm trial data in a similar manner as in Achana \emph{et al.} \cite{achana2014}.
Another limitation is related to  the prior distribution for the ancillary variance-covariance matrix in model 1b, introduced to allow for the second order consistency constraints. It was based on the Cholesky decomposition which results in the prior distributions for the between-studies correlations being dependent on the ordering of the treatments in the network. We carried out a sensitivity analysis by changing the ordering of treatments in the illustrative example in aCRC; the results remained very similar to those obtained from the main analysis.
%We would recommend that such sensitivity analysis is carried out.
Scarcity of data may also present a problem in fitting the models. For estimation of the surrogate relationships within the treatment contrasts, a relatively large number of trials per treatment contrast is needed, whereas for the across-treatment surrogacy, a range of treatments needs to be included in the network.

Nevertheless, we believe that the models have  great potential for making improvements in the research area of surrogate endpoint evaluation.
In our example in aCRC, the models allowed us to disentangle information on a relatively strong surrogate relationship between treatment effects on TR and PFS for the treatment contrast  of EGFRi with chemotherapy vs. chemotherapy alone from a set of treatments with suboptimal overall surrogacy relationship. Moreover, in  medical decision-making, where multiple comparisons of new health technologies against different comparators play an important role, NMA is a valuable tool in obtaining  average effects across all treatment contrasts in the network of treatments. In a similar way, our proposed methodology  can be used to predict the  effect of a new treatment on the final clinical outcome against any comparator in the network.
In conclusion, we developed a new meta-analytic method for surrogate endpoint evaluation that allows modelling of surrogate relationships in  greater detail.

\section{Acknowledgements}
This work was funded by the Medical Research Council, grant no.  MR/L009854/1 awarded to Sylwia Bujkiewicz.
Ian White and Rebecca Turner were supported by the Medical Research Council Unit Programme MC\_UU\_12023/21.
This research used the ALICE High Performance Computing Facility at the University of Leicester.

\bibliographystyle{plain}
\bibliography{myrefs}

\newpage
\appendix

\section{Beta distribution based prior for the correlation}
\label{appendix2}
\begin{figure}[h]
\centering
\includegraphics[scale=0.2]{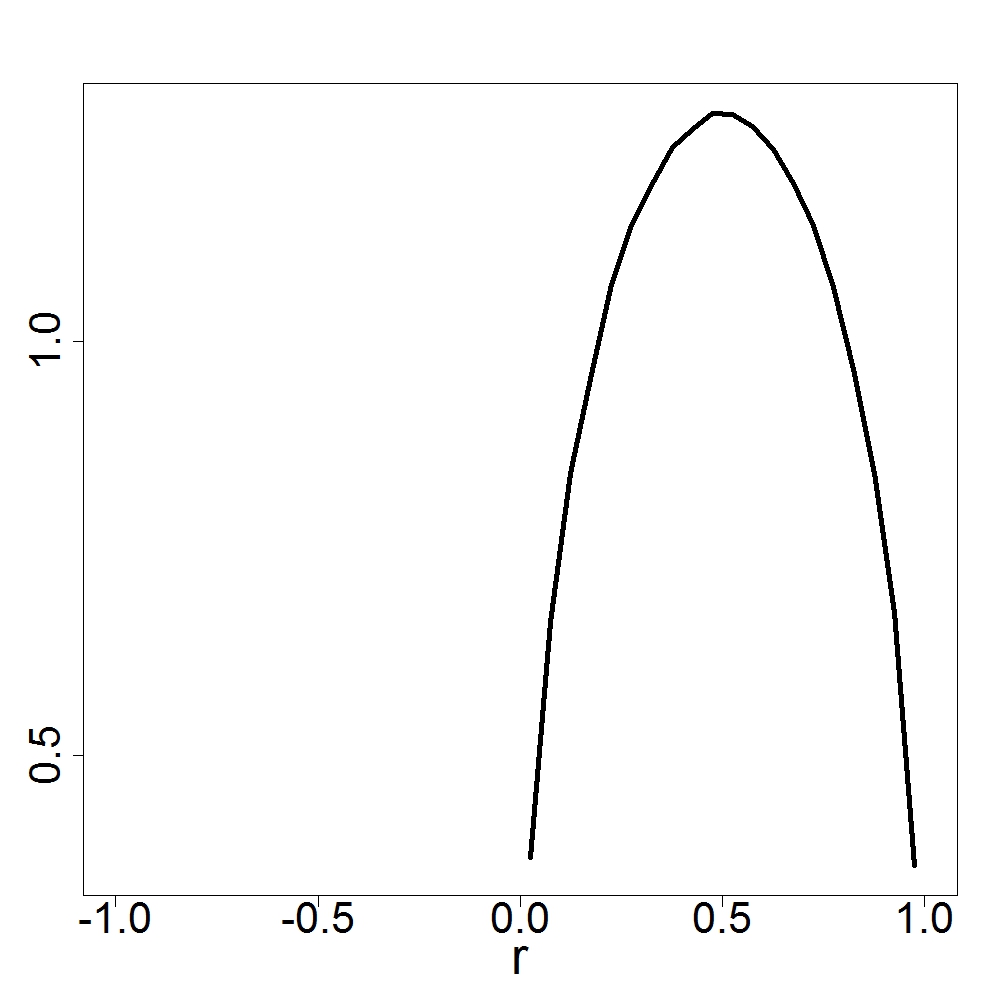}
\includegraphics[scale=0.2]{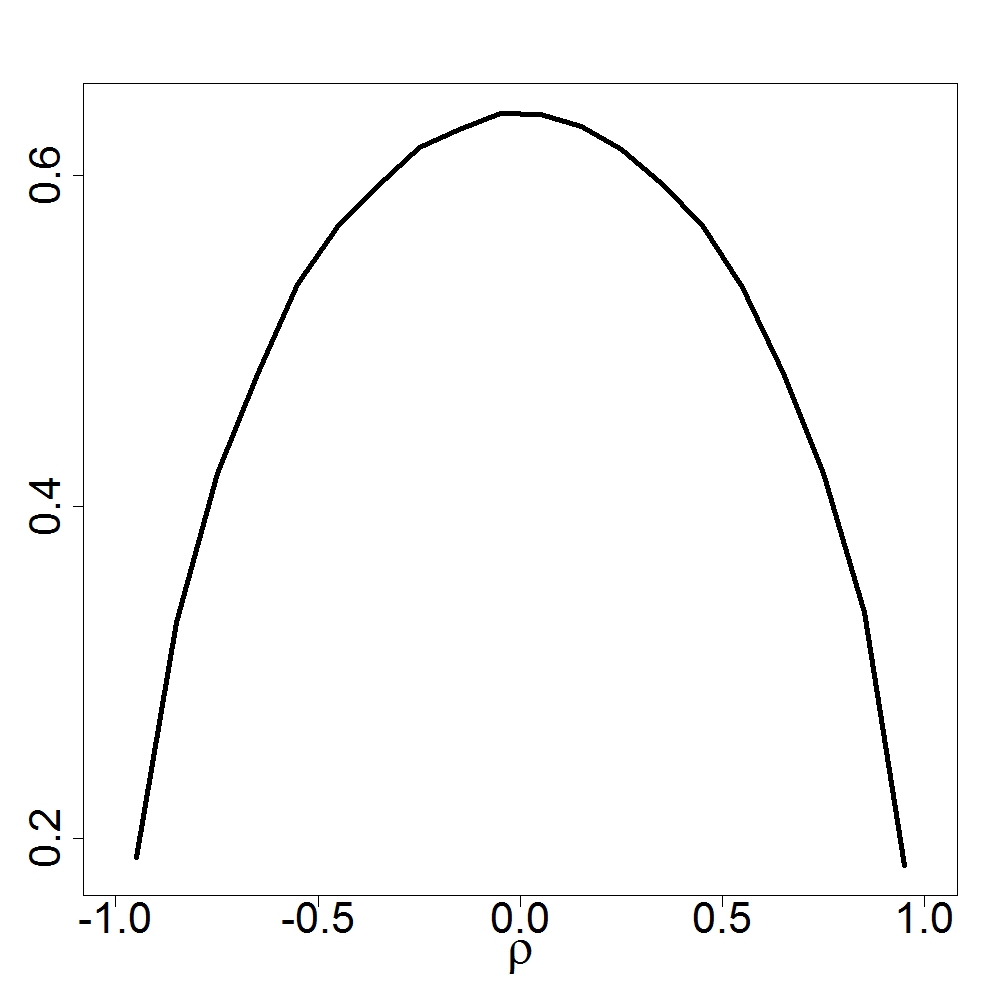}
\caption{Beta distribution
\label{fig.corr.beta}}
\end{figure}

A beta distribution was used to construct prior distributions for the between-studies correlations  (all models) and for the correlation between effects on treatment arms (models 2a--c).
A random variable drawn from a beta distribution $r \sim Beta(1.5,1.5)$ is limited to values between $0$ and $1$ with probability density zero on the edges and mean value of $0.5$, as seen in Figure \ref{fig.corr.beta} (left).
Transforming this variable, such as $\rho=r*2-1$, gives a distribution bounded by $-1$ and $1$ with mean at zero, as shown in Figure \ref{fig.corr.beta} (right). This can be used as a prior distribution for the between-studies correlation, as in Burke et al (2016).
The resulting prior distribution for the correlation, such as $\frac{\rho + 1}{2} \sim Beta(1.5,1.5)$,  allows for positive and negative values of the between-study correlation and it is relatively flat across the range of values, with the exception that values at the extreme ends of the distribution are considered extremely unlikely.

\section{Surrogacy criteria}
\label{sec.sur.crit.apx}
Daniels and Hughes defined the surrogacy criteria for a Bayesian meta-analytic model where the relationship between the true treatment effects on final clinical outcome  $\mu_{2i}$ and the effect on the surrogate endpoint $\mu_{1i}$  was written in the form of a linear regression:
\begin{equation}
\mu_{2i} \vert \mu_{1i} \sim N(\lambda_0 + \lambda_1 \mu_{1i}, \psi^2).
\end{equation}
The surrogate relationship between the two treatment effects, $\mu_{2i}$ and $\mu_{1i}$, was perfect if the intercept $\lambda_0$ was zero, as then a zero effect on a surrogate would imply a zero effect on the final outcome, the slope $\lambda_1$ should not be zero for the association to be strong, with the conditional variance $\psi^2$ being zero.
For the complete model see Daniels and Hughes (1997). A similar relationship and surrogacy criteria were described by Bujkiewicz et al (2015) in the framework of bivariate meta-analysis and extended by Bujkiewicz et al (2016) to multivariate meta-analysis. In the two papers the relationship between the regression parameters and the elements of the between-studies variance-covariance matrix was defined, similarly as in Bujkiewicz et al (2013).
The derived relationships in the bivariate case are
\begin{equation} \lambda_1=\rho_1 \frac{\tau_2}{\tau_1} \label{rel1} \end{equation}
and
\begin{equation} \psi^2=\tau_2^2 - \lambda_1^2 \tau_1^2. \label{rel2} \end{equation}

If the surrogacy relationship is perfect, the conditional variance is zero: $\psi^2=0$ (Daniels and Hughes (1997), Bujkiewicz et al (2015)).
Hence, from (\ref{rel2}), $\tau_2^2=\lambda_1^2\tau_1^2$ which gives $\lambda_1=\pm\frac{\tau_2}{\tau_1}$,
and from (\ref{rel1}) it implies that the correlation $\rho=\pm 1$.
Also $\rho^2=1$, which some authors refer to as the study level adjusted $R$-squared (Burzykowski et al (2001), Renfro et al (2012)).\\

\section{Relationships between the models}
\label{supl.models}
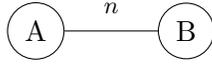
\begin{figure}[h]
%\hspace{1cm}
\centering
\begin{tikzpicture}[nodes={draw, circle}]
\centering
[inner sep=3mm]
\path
node at ( -6,-1.6) [shape=circle] (f1) {A}
node at ( -4,-1.6) [shape=circle] (f2) {B}
(f1) edge node[above, pos=0.5, draw=none] {\footnotesize $n$} (f2);
\end{tikzpicture}
\caption{Example network diagram: all $n$ studies include the same treatment contrast (only two treatments) (left).\label{fig-reduced}}
\end{figure}

The models reduce to the standard meta-analysis model for surrogate endpoints, such as the BRMA model,
%(\ref{brma-w})--(\ref{brma-b})
in a special case of data structure.
When there are only two treatments in the network, as depicted in Figure \ref{fig-reduced}, it can be shown that model 1a reduces to BRMA. Equations (\ref{eq-bvnma-w-het})--(\ref{eq-bvnma-b-het}) become
\begin{equation}
\left(
\begin{array}{c}
Y_{1(12)i}\\
Y_{2(12)i}\\
\end{array}
\right) \sim \rm{N}
\left(
\left(
\begin{array}{c}
\mu_{1(12)i}\\
\mu_{2(12)i}\\
\end{array}\right),
%\mathbf{\Sigma_i}
%\right), \;
%\mathbf{\Sigma_i}=
\left(
\begin{array}{cc}
\sigma_{1(12)i}^2 & \sigma_{1(12)i}\sigma_{2(12)i}\rho_{wi(12)}\\
\sigma_{1(12)i}\sigma_{2(12)i}\rho_{wi(12)} & \sigma_{2(12)i}^2
\end{array}
\right)
\right)
\label{eq-bvnma-w-het-2t}
\end{equation}
\begin{equation}
\left(
\begin{array}{c}
\mu_{1(12)i}\\
\mu_{2(12)i}\\
\end{array}
\right) \sim \rm{MVN}
\left(
\left(
\begin{array}{c}
d_{1(12)}\\
d_{2(12)}\\
\end{array}\right),
\;
\left(
\begin{array}{cc}
\tau_{1(12)}^2 & \tau_{1(12)}\tau_{2(12)}\rho_{(12)}\\
\tau_{1(12)}\tau_{2(12)}\rho_{(12)} & \tau_{2(12)}^2
\end{array}
\right)
\right)
\label{eq-bvnma-b-het-2t}
%\nonumber
\end{equation}
The index $(12)$ denoting the two treatments does not vary across studies or contrasts and hence can be dropped, resulting in equations for BRMA -- equations (3.1)--(3.2) in the main manuscript, with $d_j=\beta_j$ and $j=1,2$.

\section{Additional results: aCRC example}
\label{supl.results.acrc}
Tables \ref{varcov.acrc1} and \ref{varcov.acrc2} give additional resuts to the illustrative example in aCRC.
Figure \ref{acrc2.pred} shows the predicted effects obtained from BRMA and model 2d along with the observed estimates of the effects on PFS. The improvement in predictions was not substantial due to the weak association patterns between the treatment effects on the two outcomes.

\begin{table}[b]
\centering
\begin{tabular}{lcccc}
model & AB & AC & BC & BD  \\ \hline
\hline
\multicolumn{5}{l}{\emph{log OR (TR)}}  \\
BRMA&\multicolumn{4}{c}{0.54 (0.36, 0.72)}\\
model 1a&0.49 (0.26, 0.73)&0.81 (0.57, 1.08)&0.33 (0.09, 0.57)&0.16 (-0.67, 0.95)\\
model 1b&0.47 (0.22, 0.74)&0.79 (0.53, 1.07)&0.31 (0.03, 0.6)&0.15 (-0.73, 0.95)\\
model 1c&0.46 (0.24, 0.69)&0.77 (0.55, 1)&0.31 (0.06, 0.55)&0.19 (-0.28, 0.63)\\
model 1d&0.49 (0.21, 0.77)&0.74 (0.48, 1)&0.25 (-0.09, 0.58)&0.18 (-0.39, 0.75)\\
model 2a&0.45 (0.22, 0.69)&0.76 (0.49, 1.02)&0.3 (0.06, 0.53)&0.11 (-0.45, 0.65)\\
model 2b&0.44 (0.18, 0.69)&0.72 (0.45, 0.99)&0.28 (0.01, 0.55)&0.1 (-0.46, 0.63)\\
model 2c&0.43 (0.21, 0.65)&0.72 (0.49, 0.94)&0.29 (0.06, 0.51)&0.15 (-0.23, 0.52)\\
model 2d&0.45 (0.18, 0.72)&0.68 (0.44, 0.93)&0.23 (-0.07, 0.54)&0.12 (-0.34, 0.6)\\
\hline
\multicolumn{5}{l}{\emph{log HR (PFS)}} \\
BRMA&\multicolumn{4}{c}{-0.24 (-0.33, -0.16)}\\
model 1a&-0.37 (-0.47, -0.27)&-0.31 (-0.42, -0.2)&0.06 (-0.05, 0.18)&0.09 (-0.33, 0.51)\\
model 1b&-0.37 (-0.48, -0.26)&-0.3 (-0.41, -0.18)&0.07 (-0.05, 0.21)&0.09 (-0.28, 0.49)\\
model 1c&-0.37 (-0.46, -0.27)&-0.3 (-0.4, -0.2)&0.07 (-0.04, 0.19)&0.08 (-0.14, 0.3)\\
model 1d&-0.38 (-0.51, -0.26)&-0.28 (-0.39, -0.18)&0.1 (-0.05, 0.25)&0.11 (-0.15, 0.37)\\
model 2a&-0.35 (-0.45, -0.24)&-0.29 (-0.4, -0.18)&0.06 (-0.05, 0.17)&0.1 (-0.16, 0.37)\\
model 2b&-0.34 (-0.45, -0.23)&-0.28 (-0.39, -0.17)&0.07 (-0.05, 0.2)&0.1 (-0.14, 0.36)\\
model 2c&-0.34 (-0.44, -0.24)&-0.28 (-0.38, -0.18)&0.06 (-0.04, 0.17)&0.09 (-0.09, 0.28)\\
model 2d&-0.35 (-0.47, -0.23)&-0.27 (-0.37, -0.17)&0.08 (-0.05, 0.22)&0.13 (-0.09, 0.34)\\
\hline
\multicolumn{5}{l}{\emph{Var(log OR) (TR)}}  \\
BRMA&\multicolumn{4}{c}{0.3 (0.16, 0.52)} \\
model 1a&0.23 (0.08, 0.59)&0.58 (0.23, 1.26)&0.11 (0, 0.8)&0.87 (0.03, 3.32) \\
model 1b&0.31 (0.11, 0.7)&0.45 (0.2, 0.88)&0.13 (0, 0.58)&0.71 (0.04, 3.01) \\
model 1c&0.24 (0.09, 0.49)&0.32 (0.15, 0.62)&0.07 (0, 0.32)&0.15 (0.01, 0.52) \\
model 1d&\multicolumn{4}{c}{0.3 (0.15, 0.55)} \\
model 2a&0.23 (0.07, 0.56)&0.59 (0.23, 1.31)&0.12 (0, 0.85)&0.76 (0.02, 3.16) \\
model 2b&0.31 (0.11, 0.68)&0.44 (0.2, 0.86)&0.13 (0, 0.56)&0.6 (0.03, 2.66) \\
model 2c&0.23 (0.09, 0.45)&0.3 (0.15, 0.56)&0.06 (0, 0.23)&0.13 (0.01, 0.5) \\
model 2d&\multicolumn{4}{c}{0.29 (0.15, 0.53)} \\
\hline
\multicolumn{5}{l}{\emph{Var(log HR) (PFS)}} \\
BRMA&\multicolumn{4}{c}{0.08 (0.04, 0.13)} \\
model 1a&0.03 (0.01, 0.09)&0.11 (0.05, 0.21)&0.04 (0, 0.3)&0.24 (0, 1.47) \\
model 1b&0.05 (0.01, 0.12)&0.09 (0.04, 0.16)&0.03 (0, 0.12)&0.16 (0.01, 0.92)  \\
model 1c&0.04 (0.01, 0.08)&0.06 (0.03, 0.11)&0.02 (0, 0.07)&0.03 (0, 0.13)  \\
model 1d&\multicolumn{4}{c}{0.06 (0.03, 0.1)}  \\
model 2a&0.03 (0.01, 0.1)&0.11 (0.05, 0.22)&0.03 (0, 0.2)&0.18 (0.01, 1.03)  \\
model 2b&0.05 (0.01, 0.12)&0.08 (0.04, 0.16)&0.03 (0, 0.12)&0.12 (0.01, 0.61)  \\
model 2c&0.04 (0.01, 0.08)&0.06 (0.03, 0.11)&0.02 (0, 0.06)&0.03 (0, 0.11)  \\
model 2d&\multicolumn{4}{c}{0.06 (0.03, 0.1)}  \\
\hline
\end{tabular}
\caption{Mean effects and the between-studies variances for each model in the aCRC example.
A -- chemotherapy alone, B -- anti-VEGF therapies + chemotherapy,
C -- EGFRi  + chemotherapy, D -- EGFRi + anti-VEGF therapies + chemotherapy.}
\label{varcov.acrc1}
\end{table}

\clearpage
%\newpage

\begin{table}[t]
\centering
\begin{tabular}{lccc}
model & AE & AD & CF \\ \hline
\multicolumn{4}{l}{\emph{log OR (TR)}}\\
model 1a&1.4 (-1.86, 4.66)&0.64 (-0.2, 1.46)&-0.01 (-1.49, 1.49)\\
model 1b&1.41 (-2.02, 4.82)&0.62 (-0.28, 1.45)&-0.01 (-1.56, 1.54)\\
model 1c&1.39 (-1.08, 3.87)&0.65 (0.14, 1.14)&-0.02 (-0.7, 0.66)\\
model 1d&1.42 (-1.06, 3.88)&0.67 (0.05, 1.28)&-0.02 (-0.92, 0.88)\\
model 2a&0.57 (-0.38, 1.75)&0.56 (-0.01, 1.14)&-0.16 (-0.85, 0.5)\\
model 2b&0.54 (-0.45, 1.74)&0.54 (-0.04, 1.1)&-0.16 (-0.87, 0.53)\\
model 2c&0.57 (-0.3, 1.7)&0.58 (0.16, 0.99)&-0.12 (-0.6, 0.39)\\
model 2d&0.57 (-0.32, 1.75)&0.57 (0.08, 1.08)&-0.12 (-0.72, 0.52)\\
\hline
\multicolumn{4}{l}{\emph{log HR (PFS)}}\\
model 1a&0 (-2.55, 2.55)&-0.28 (-0.7, 0.15)&0.12 (-1.21, 1.46)\\
model 1b&0.01 (-2.58, 2.61)&-0.27 (-0.66, 0.13)&0.11 (-1.24, 1.47)\\
model 1c&0 (-0.7, 0.68)&-0.28 (-0.52, -0.05)&0.11 (-0.24, 0.47)\\
model 1d&0 (-0.69, 0.69)&-0.27 (-0.55, 0.01)&0.11 (-0.31, 0.54)\\
model 2a&-0.19 (-0.71, 0.32)&-0.24 (-0.52, 0.03)&0.08 (-0.3, 0.45)\\
model 2b&-0.19 (-0.69, 0.33)&-0.24 (-0.5, 0.02)&0.07 (-0.31, 0.45)\\
model 2c&-0.14 (-0.51, 0.28)&-0.25 (-0.45, -0.05)&0.09 (-0.17, 0.35)\\
model 2d&-0.13 (-0.5, 0.31)&-0.23 (-0.46, -0.01)&0.1 (-0.19, 0.4)\\
\hline
\multicolumn{4}{l}{\emph{correlations}}\\
model 1a&0 (-0.88, 0.88)&0 (-0.89, 0.89)&0.01 (-0.9, 0.9)\\
model 1b&-0.08 (-0.89, 0.83)&-0.32 (-0.95, 0.72)&-0.02 (-0.84, 0.82)\\
model 1c&-0.48 (-0.96, 0.64)&-0.46 (-0.96, 0.64)&-0.15 (-0.87, 0.75)\\
model 2a&0.02 (-0.87, 0.89)&-0.01 (-0.89, 0.88)&0 (-0.9, 0.9)\\
model 2b&-0.07 (-0.89, 0.84)&-0.35 (-0.95, 0.7)&-0.04 (-0.85, 0.82)\\
model 2c&-0.5 (-0.96, 0.57)&-0.49 (-0.96, 0.56)&-0.14 (-0.86, 0.75)\\
\hline
\multicolumn{4}{l}{\emph{Var(log OR) (TR)}}\\
model 1a&1.33 (0, 3.81)&0.98 (0, 3.65)&0.79 (0, 3.49)\\
model 1b&1.67 (0.09, 5.19)&0.89 (0.04, 3.51)&0.92 (0.01, 3.74)\\
model 1c&0.31 (0.02, 1.07)&0.29 (0.02, 0.92)&0.14 (0, 0.62)\\
model 2a&1.21 (0, 3.75)&0.95 (0, 3.64)&0.63 (0, 3.31)\\
model 2b&1.5 (0.08, 4.92)&0.78 (0.03, 3.13)&0.7 (0.01, 3.42)\\
model 2c&0.28 (0.03, 0.96)&0.27 (0.03, 0.84)&0.1 (0, 0.46)\\
\hline
\multicolumn{4}{l}{\emph{Var(log HR) (PFS)}}\\
model 1a&1.34 (0, 3.81)&0.81 (0, 3.56)&0.67 (0, 3.49)\\
model 1b&1.38 (0.02, 4.01)&0.19 (0.01, 0.97)&0.72 (0, 3.51)\\
model 1c&0.06 (0, 0.22)&0.05 (0, 0.18)&0.03 (0, 0.14)\\
model 2a&0.85 (0, 3.57)&0.79 (0, 3.56)&0.36 (0, 2.7)\\
model 2b&0.91 (0.01, 3.69)&0.15 (0.01, 0.67)&0.39 (0, 2.75)\\
model 2c&0.05 (0, 0.19)&0.05 (0, 0.17)&0.03 (0, 0.12)\\
\hline
\end{tabular}
\caption{Mean effects and the between-studies correlations and variances for each model in the aCRC example (contrasts AE, AD and CF).
A -- chemotherapy alone,
C -- EGFRi  + chemotherapy, D -- EGFRi + anti-VEGF therapies + chemotherapy, E -- anti-IGF1R , F -- anti-IgG2 + chemotherapy}
\label{varcov.acrc2}
\end{table}

\clearpage

%\begin{figure}[b]
%centering
%\includegraphics[scale=0.15]{ResultsACRC2_newdata_m2_Dec2017_brma_2d_sorted2.jpg}
%\caption{Predicted effects obtained from BRMA and model 2d along with the observed estimates of the effects on PFS for aCRC data}
%\label{acrc2.pred}
%\end{figure}

%\clearpage

%\newpage
\section{Sensitivity analysis (aCRC example)}
\label{sec.sens.anal}
\begin{figure}[h]
\label{scat.acrc.sens}
\centering
\includegraphics[scale=0.3]{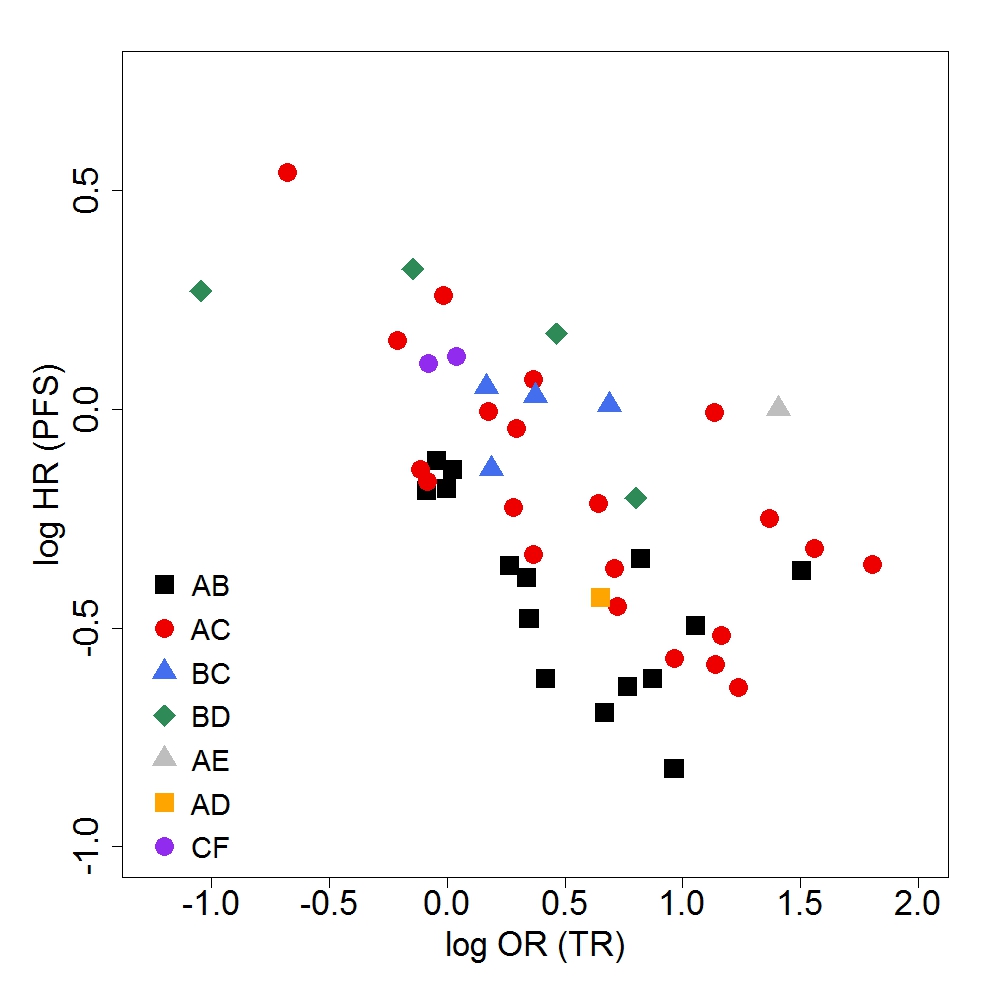}
\begin{tikzpicture}[nodes={draw, circle}, line width=1.2pt]
[inner sep=3mm]
\path node at ( -3,4) [shape=circle,draw=black!40] (b) {B}
node at ( 0,4) [shape=circle,draw=black!40] (a) {A}
node at ( -3,1) [shape=circle,draw=black!40] (d) {D}
node at ( 0,1) [shape=circle,draw=black!40] (c) {C}
node at ( 3,1) [shape=circle,draw=black!40] (f) {F}
node at ( 3,4) [shape=circle,draw=black!40] (e) {E}
node at (0,-1) [shape=circle,draw=white] (ghost) {}
(a) edge[black] node[above, align=flush center, draw=none, xshift= 2pt]
                                {\footnotesize \textcolor[rgb]{0.00,0.00,0.00}{$15$}} (b)
(a) edge[red] node[right, align=flush center, draw=none, xshift= 2pt]
                                {\footnotesize \textcolor[rgb]{0.00,0.00,0.00}{$21$}} (c)
(b) edge[blue!70!cyan] node[below, align=flush center, draw=none, xshift= 6pt, yshift= -8pt]
                                    {\footnotesize \textcolor[rgb]{0.00,0.00,0.00}{$4$}} (c)
(b) edge[teal] node[right, align=flush center, draw=none, xshift= -2pt] {\footnotesize \textcolor[rgb]{0.00,0.00,0.00}{$4$}} (d)
(a) edge[gray!80] node[above, align=flush center, draw=none] {\footnotesize \textcolor[rgb]{0.00,0.00,0.00}{$1$}} (e)
(a) edge[orange] node[above, align=flush center, draw=none, xshift= 7pt, yshift= 7pt] {\footnotesize \textcolor[rgb]{0.00,0.00,0.00}{$1$}} (d)
(c) edge[violet!80] node[above, align=flush center, draw=none] {\footnotesize \textcolor[rgb]{0.00,0.00,0.00}{$2$}} (f);
\end{tikzpicture}
\caption{Scatter plot and network diagram for the advanced colorectal cancer example, A -- chemotherapy alone, B -- anti-VEGF therapies + chemotherapy,
C -- EGFRi  + chemotherapy, D -- EGFRi + anti-VEGF therapies + chemotherapy, E -- anti-IGF1R , F -- anti-IgG2 + chemotherapy}
\end{figure}
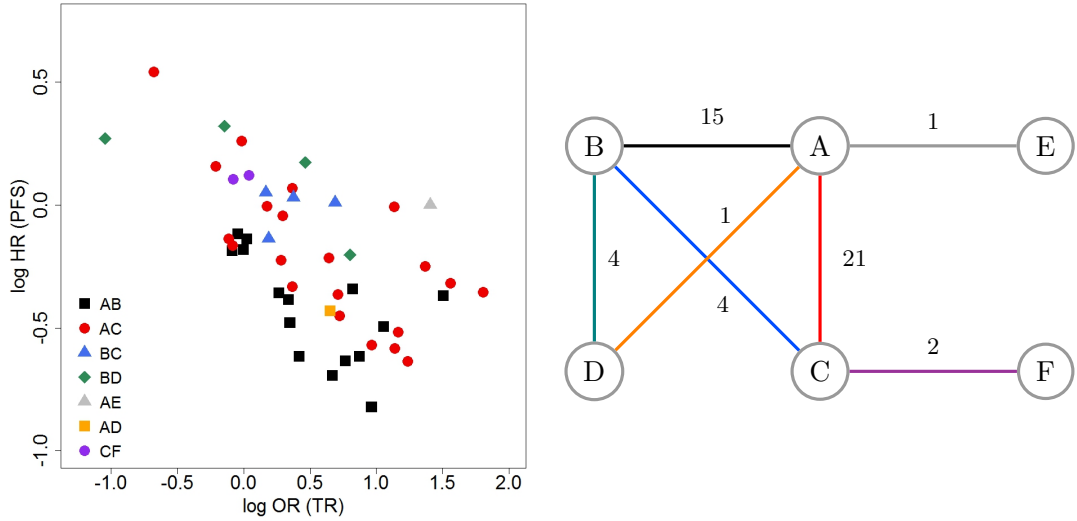

Sensitivity analysis was carried out investigating the effect of potentially influential observations (three studies with largest treatment effect on TR, due to no events in the control arm, were removed). Figure \ref{scat.acrc.sens} shows the scatter plot. Tables \ref{sens1} and \ref{sens2} show the between studies correlations of the heterogeneity parameters.

%\newpage
\begin{table}[p]
\centering
\begin{tabular}{lcccc}
model & AB & BC & AC & BD \\ \hline
\multicolumn{5}{l}{\emph{log OR (TR)}}\\
BRMA&\multicolumn{4}{c}{0.48 (0.32, 0.65)}\\
model 1a&0.43 (0.2, 0.67)&0.71 (0.46, 0.96)&0.28 (0, 0.52)&0.16 (-0.69, 0.95)\\
model 1b&0.42 (0.17, 0.67)&0.68 (0.43, 0.93)&0.26 (-0.04, 0.52)&0.16 (-0.7, 0.95)\\
model 1c&0.41 (0.19, 0.64)&0.68 (0.45, 0.9)&0.27 (0, 0.5)&0.19 (-0.3, 0.64)\\
model 1d&0.46 (0.2, 0.72)&0.63 (0.39, 0.86)&0.17 (-0.15, 0.48)&0.18 (-0.36, 0.72)\\
model 2a&0.39 (0.15, 0.63)&0.65 (0.39, 0.9)&0.26 (-0.01, 0.48)&0.11 (-0.43, 0.66)\\
model 2b&0.38 (0.13, 0.64)&0.62 (0.36, 0.87)&0.24 (-0.05, 0.49)&0.1 (-0.44, 0.62)\\
model 2c&0.37 (0.15, 0.6)&0.63 (0.4, 0.85)&0.25 (0.02, 0.47)&0.15 (-0.24, 0.51)\\
model 2d&0.42 (0.17, 0.67)&0.58 (0.34, 0.81)&0.16 (-0.13, 0.45)&0.1 (-0.33, 0.56)\\
\hline
\multicolumn{5}{l}{\emph{log HR (PFS)}}\\
BRMA&\multicolumn{4}{c}{-0.21 (-0.29, -0.12)}\\
model 1a&-0.35 (-0.45, -0.24)&-0.24 (-0.36, -0.13)&0.11 (-0.01, 0.26)&0.09 (-0.35, 0.52)\\
model 1b&-0.35 (-0.45, -0.24)&-0.23 (-0.34, -0.13)&0.11 (-0.01, 0.25)&0.08 (-0.3, 0.47)\\
model 1c&-0.34 (-0.44, -0.25)&-0.23 (-0.33, -0.14)&0.11 (0.01, 0.23)&0.08 (-0.13, 0.3)\\
model 1d&-0.36 (-0.47, -0.25)&-0.21 (-0.31, -0.12)&0.15 (0.02, 0.28)&0.1 (-0.12, 0.33)\\
model 2a&-0.32 (-0.43, -0.21)&-0.23 (-0.34, -0.12)&0.09 (-0.02, 0.23)&0.11 (-0.15, 0.37)\\
model 2b&-0.32 (-0.43, -0.2)&-0.22 (-0.32, -0.12)&0.1 (-0.01, 0.23)&0.1 (-0.14, 0.35)\\
model 2c&-0.32 (-0.41, -0.21)&-0.22 (-0.31, -0.13)&0.09 (-0.01, 0.2)&0.09 (-0.08, 0.26)\\
model 2d&-0.33 (-0.44, -0.23)&-0.2 (-0.3, -0.11)&0.13 (0.01, 0.25)&0.12 (-0.07, 0.31)\\
\hline
\multicolumn{5}{l}{\emph{correlations}}\\
BRMA&\multicolumn{4}{c}{-0.67 (-0.87, -0.4)}\\
model 1a&-0.53 (-0.91, 0.05)&-0.74 (-0.96, -0.34)&-0.03 (-0.89, 0.87)&-0.28 (-0.94, 0.65)\\
model 1b&-0.66 (-0.95, -0.13)&-0.76 (-0.96, -0.41)&-0.21 (-0.89, 0.71)&-0.28 (-0.9, 0.58)\\
model 1c&-0.66 (-0.95, -0.16)&-0.76 (-0.97, -0.4)&-0.19 (-0.87, 0.7)&-0.32 (-0.89, 0.51)\\
model 1d&\multicolumn{4}{c}{-0.76 (-0.94, -0.47)}\\
model 2a&-0.55 (-0.92, 0.03)&-0.73 (-0.95, -0.32)&-0.03 (-0.9, 0.88)&-0.3 (-0.95, 0.65)\\
model 2b&-0.67 (-0.95, -0.15)&-0.77 (-0.96, -0.41)&-0.2 (-0.89, 0.72)&-0.3 (-0.89, 0.55)\\
model 2c&-0.71 (-0.96, -0.27)&-0.81 (-0.97, -0.52)&-0.25 (-0.89, 0.68)&-0.3 (-0.89, 0.54)\\
model 2d&\multicolumn{4}{c}{-0.76 (-0.95, -0.49)}\\
\hline
\multicolumn{5}{l}{\emph{Var(log OR) (TR)}}\\
BRMA&\multicolumn{4}{c}{0.26 (0.14, 0.43)}\\
model 1a&0.23 (0.07, 0.59)&0.41 (0.17, 0.89)&0.13 (0, 0.91)&0.86 (0.03, 3.33)\\
model 1b&0.27 (0.1, 0.58)&0.34 (0.16, 0.66)&0.12 (0, 0.56)&0.69 (0.04, 2.99)\\
model 1c&0.21 (0.08, 0.44)&0.26 (0.12, 0.49)&0.07 (0, 0.32)&0.16 (0.01, 0.57)\\
model 1d&\multicolumn{4}{c}{0.24 (0.13, 0.43)}\\
model 2a&0.25 (0.08, 0.61)&0.4 (0.17, 0.83)&0.13 (0, 0.9)&0.75 (0.02, 3.1)\\
model 2b&0.27 (0.1, 0.58)&0.34 (0.16, 0.66)&0.12 (0, 0.52)&0.59 (0.03, 2.58)\\
model 2c&0.23 (0.09, 0.45)&0.3 (0.15, 0.56)&0.06 (0, 0.23)&0.13 (0.01, 0.5)\\
model 2d&\multicolumn{4}{c}{0.24 (0.13, 0.4)}\\
\hline
\multicolumn{5}{l}{\emph{Var(log HR) (PFS)}}\\
BRMA&\multicolumn{4}{c}{0.06 (0.03, 0.1)}\\
model 1a&0.03 (0.01, 0.1)&0.06 (0.02, 0.15)&0.06 (0, 0.42)&0.26 (0.01, 1.71)\\
model 1b&0.04 (0.01, 0.1)&0.05 (0.02, 0.11)&0.03 (0, 0.1)&0.16 (0.01, 0.88)\\
model 1c&0.03 (0.01, 0.07)&0.04 (0.01, 0.08)&0.02 (0, 0.06)&0.03 (0, 0.09)\\
model 1d&\multicolumn{4}{c}{0.04 (0.02, 0.07)}\\
model 2a&0.04 (0.01, 0.11)&0.06 (0.02, 0.14)&0.05 (0, 0.32)&0.18 (0.01, 1.02)\\
model 2b&0.04 (0.01, 0.1)&0.05 (0.02, 0.11)&0.02 (0, 0.09)&0.12 (0.01, 0.62)\\
model 2c&0.04 (0.01, 0.08)&0.06 (0.03, 0.11)&0.02 (0, 0.06)&0.03 (0, 0.11)\\
model 2d&\multicolumn{4}{c}{0.04 (0.02, 0.07)}\\
\hline
\end{tabular}
\caption{Between-studies correlations for each model in the aCRC example. Where only one value is given (models BRMA, 1d and 2d), the parameters are common across the treatment contrasts.}
\label{sens1}
\end{table}

\newpage
\begin{table}[p]
\centering
\begin{tabular}{lccc}
model & AE & AD & CF \\ \hline
\multicolumn{4}{l}{\emph{log OR (TR)}}\\
model 1a&1.41 (-1.82, 4.64)&0.59 (-0.27, 1.39)&-0.02 (-1.55, 1.52)\\
model 1b&1.42 (-1.95, 4.79)&0.58 (-0.3, 1.4)&-0.02 (-1.55, 1.5)\\
model 1c&1.4 (-1.04, 3.86)&0.61 (0.09, 1.1)&-0.02 (-0.67, 0.63)\\
model 1d&1.44 (-1.01, 3.88)&0.64 (0.06, 1.21)&-0.02 (-0.85, 0.8)\\
model 2a&0.52 (-0.41, 1.72)&0.51 (-0.06, 1.08)&-0.13 (-0.8, 0.52)\\
model 2b&0.5 (-0.41, 1.68)&0.49 (-0.07, 1.05)&-0.14 (-0.81, 0.53)\\
model 2c&0.51 (-0.29, 1.55)&0.52 (0.1, 0.93)&-0.11 (-0.57, 0.37)\\
model 2d&0.5 (-0.33, 1.62)&0.52 (0.07, 1.02)&-0.1 (-0.66, 0.5)\\
\hline
\multicolumn{4}{l}{\emph{log HR (PFS)}}\\
model 1a&0.01 (-2.55, 2.53)&-0.26 (-0.7, 0.18)&0.11 (-1.18, 1.38)\\
model 1b&0 (-2.57, 2.58)&-0.26 (-0.65, 0.13)&0.11 (-1.25, 1.48)\\
model 1c&-0.01 (-0.66, 0.64)&-0.26 (-0.49, -0.03)&0.12 (-0.22, 0.46)\\
model 1d&0 (-0.64, 0.63)&-0.26 (-0.5, -0.02)&0.11 (-0.25, 0.48)\\
model 2a&-0.16 (-0.65, 0.33)&-0.22 (-0.5, 0.05)&0.06 (-0.29, 0.41)\\
model 2b&-0.17 (-0.66, 0.32)&-0.22 (-0.47, 0.04)&0.06 (-0.31, 0.41)\\
model 2c&-0.13 (-0.46, 0.24)&-0.23 (-0.41, -0.04)&0.07 (-0.15, 0.31)\\
model 2d&-0.11 (-0.46, 0.29)&-0.22 (-0.42, -0.02)&0.08 (-0.18, 0.35)\\
\hline
\multicolumn{4}{l}{\emph{correlations}}\\
model 1a&0 (-0.88, 0.88)&-0.01 (-0.89, 0.89)&0.02 (-0.88, 0.9)\\
model 1b&-0.06 (-0.87, 0.83)&-0.27 (-0.94, 0.75)&-0.02 (-0.83, 0.82)\\
model 1c&-0.42 (-0.96, 0.69)&-0.38 (-0.95, 0.69)&-0.11 (-0.85, 0.77)\\
model 2a&0.02 (-0.88, 0.89)&-0.02 (-0.89, 0.88)&-0.01 (-0.91, 0.89)\\
model 2b&-0.06 (-0.88, 0.84)&-0.32 (-0.94, 0.72)&-0.02 (-0.83, 0.81)\\
model 2c&-0.5 (-0.96, 0.57)&-0.49 (-0.96, 0.56)&-0.14 (-0.86, 0.75)\\
\hline
\multicolumn{4}{l}{\emph{Var(log OR) (TR)}}\\
model 1a&1.34 (0, 3.8)&0.99 (0, 3.67)&0.85 (0, 3.6)\\
model 1b&1.6 (0.07, 4.93)&0.81 (0.03, 3.32)&0.91 (0.01, 3.72)\\
model 1c&0.26 (0.02, 0.96)&0.25 (0.02, 0.9)&0.11 (0, 0.51)\\
model 2a&1.19 (0, 3.74)&0.94 (0, 3.64)&0.58 (0, 3.19)\\
model 2b&1.45 (0.06, 4.73)&0.72 (0.03, 2.94)&0.66 (0.01, 3.33)\\
model 2c&0.28 (0.03, 0.96)&0.27 (0.03, 0.84)&0.1 (0, 0.46)\\
\hline
\multicolumn{4}{l}{\emph{Var(log HR) (PFS)}}\\
model 1a&1.33 (0, 3.8)&0.81 (0, 3.57)&0.63 (0, 3.39)\\
model 1b&1.37 (0.02, 3.93)&0.17 (0, 0.91)&0.7 (0, 3.48)\\
model 1c&0.04 (0, 0.17)&0.04 (0, 0.13)&0.03 (0, 0.11)\\
model 2a&0.83 (0, 3.57)&0.8 (0, 3.54)&0.33 (0, 2.55)\\
model 2b&0.87 (0.01, 3.63)&0.14 (0.01, 0.66)&0.36 (0, 2.63)\\
model 2c&0.05 (0, 0.19)&0.05 (0, 0.17)&0.03 (0, 0.12)\\
\hline
\end{tabular}
\caption{Between-studies correlations and variances  for each model in the aCRC example (contrasts AE, AD and CF).}
\label{sens2}
\end{table}

\clearpage
%\newpage

\section{Additional results: simulated scenario 1}
\label{supl.sim1}
\begin{table}[h]
\centering
\begin{tabular}{lccc}
model & AB & BC & AC \\ \hline
\multicolumn{4}{l}{\emph{surrogate endpoint $Y_1$}}\\
BRMA&\multicolumn{3}{c}{0.91 (0.52, 1.56)}\\
model 1a&0.13 (0.02, 0.4)&0.25 (0.07, 0.67)&0.47 (0.17, 1.15)\\
model 1b&0.15 (0.03, 0.44)&0.23 (0.08, 0.52)&0.46 (0.19, 0.94)\\
model 1c&0.10 (0.03, 0.25)&0.16 (0.06, 0.34)&0.32 (0.15, 0.61)\\
model 1d&\multicolumn{3}{c}{0.25 (0.13, 0.45)}\\
model 2a&0.13 (0.02, 0.39)&0.26 (0.08, 0.70)&0.48 (0.16, 1.23)\\
model 2b&0.16 (0.03, 0.45)&0.23 (0.08, 0.53)&0.46 (0.18, 0.94)\\
model 2c&0.09 (0.02, 0.24)&0.16 (0.06, 0.35)&0.33 (0.15, 0.61)\\
model 2d&\multicolumn{3}{c}{0.25 (0.13, 0.44)}\\
\hline
\multicolumn{4}{l}{\emph{final outcome $Y_2$}}\\
BRMA &\multicolumn{3}{c}{0.93 (0.53, 1.6)}\\
model 1a &0.4 (0.12, 1.11)&0.09 (0.01, 0.3)&0.4 (0.14, 1.00)\\
model 1b &0.33 (0.13, 0.72)&0.10 (0.02, 0.31)&0.45 (0.17, 0.93)\\
model 1c &0.22 (0.10, 0.45)&0.07 (0.01, 0.19)&0.31 (0.14, 0.59)\\
model 1d &\multicolumn{3}{c}{0.26 (0.13, 0.47)}\\
model 2a &0.41 (0.13, 1.10)&0.09 (0.01, 0.30)&0.41 (0.13, 1.10)\\
model 2b &0.33 (0.13, 0.71)&0.1 (0.02, 0.32)&0.45 (0.17, 0.94)\\
model 2c &0.22 (0.10, 0.45)&0.06 (0.01, 0.18)&0.31 (0.13, 0.59)\\
model 2d &\multicolumn{3}{c}{0.25 (0.13, 0.47)}\\ \hline
\end{tabular}
\caption{Heterogeneity parameters for each model under simulation scenario 1. Where only one value is given (models BRMA, 1d and 2d), the parameters are common across the treatment contrasts.}
\label{varcov.sim1}
\end{table}

\clearpage

%\newpage
\section{Additional results: simulated scenario 2}
\label{supl.sim2}
\begin{table}[h]
\centering
\begin{tabular}{lccc}
model & AB & BC & AC \\ \hline
\multicolumn{4}{l}{\emph{surrogate endpoint $Y_1$}}\\
BRMA&\multicolumn{3}{c}{0.85 (0.50, 1.42)}\\
model 1a&0.04 (0.01, 0.11)&0.12 (0.04, 0.34)&0.11 (0.04, 0.26)\\
model 1b&0.05 (0.01, 0.14)&0.08 (0.03, 0.17)&0.14 (0.05, 0.28)\\
model 1c&0.03 (0.01, 0.09)&0.06 (0.03, 0.12)&0.1 (0.05, 0.20)\\
model 1d&\multicolumn{3}{c}{0.08 (0.04, 0.14)}\\
model 2a&0.04 (0.01, 0.11)&0.12 (0.04, 0.34)&0.11 (0.04, 0.28)\\
model 2b&0.05 (0.01, 0.13)&0.08 (0.03, 0.18)&0.14 (0.05, 0.28)\\
model 2c&0.03 (0.01, 0.08)&0.06 (0.03, 0.12)&0.10 (0.05, 0.20)\\
model 2d&\multicolumn{3}{c}{0.08 (0.04, 0.14)}\\
\hline
\multicolumn{4}{l}{\emph{final outcome $Y_2$}}\\
BRMA&\multicolumn{3}{c}{0.81 (0.48, 1.36)}\\
model 1a&0.08 (0.02, 0.21)&0.12 (0.04, 0.33)&0.04 (0.01, 0.12)\\
model 1b&0.12 (0.04, 0.27)&0.07 (0.03, 0.16)&0.05 (0.02, 0.12)\\
model 1c&0.09 (0.03, 0.19)&0.05 (0.02, 0.11)&0.04 (0.01, 0.08)\\
model 1d&\multicolumn{3}{c}{0.07 (0.03, 0.12)}\\
model 2a&0.08 (0.02, 0.21)&0.12 (0.04, 0.35)&0.05 (0.01, 0.12)\\
model 2b&0.12 (0.04, 0.26)&0.07 (0.03, 0.16)&0.05 (0.02, 0.12)\\
model 2c&0.09 (0.03, 0.18)&0.05 (0.02, 0.11)&0.04 (0.01, 0.09)\\
model 2d&\multicolumn{3}{c}{0.07 (0.03, 0.12)}\\
\hline
\end{tabular}
\caption{Heterogeneity parameters for each model under simulation scenario 2. Where only one value is given (models BRMA, 1d and 2d), the parameters are common across the treatment contrasts.}
\label{varcov.sim2}
\end{table}

\begin{table}[h]
\centering
\begin{tabular}{lccccc}
&   $p_{overlap}$  &  $\vert m_{obs}-m_{pred} \vert$  &  $w_{pred}/w_{obs}$ & $\pi$ & $\% red.$ \\ \hline
\hline
\multicolumn{6}{l}{AB}\\
BRMA&1&0.13&3.35&0.31&\\
model 1a&0.99&0.13&2.37&0.43&28.61\\
model 1b&1&0.11&2.81&0.37&15.44\\
model 1c&1&0.12&2.59&0.4&21.99\\
model 1d&0.98&0.19&2.63&0.39&21.53\\
model 2a&0.99&0.13&2.35&0.43&29.02\\
model 2b&1&0.11&2.82&0.37&15.23\\
model 2c&1&0.12&2.59&0.4&21.97\\
model 2d&0.98&0.19&2.63&0.39&21.63\\
\hline
\multicolumn{6}{l}{BC}\\
BRMA&0.69&0.37&3.53&0.21&\\
model 1a&0.95&0.29&4.23&0.24&-17.64\\
model 1b&0.86&0.3&3.31&0.28&7.33\\
model 1c&0.83&0.3&2.94&0.3&17.79\\
model 1d&0.71&0.31&2.93&0.26&17.28\\
model 2a&0.95&0.29&4.22&0.24&-17.53\\
model 2b&0.86&0.3&3.31&0.28&7.46\\
model 2c&0.83&0.3&2.93&0.3&17.98\\
model 2d&0.71&0.31&2.92&0.26&17.47\\
\hline
\multicolumn{6}{l}{AC}\\
BRMA&1&0.09&3.77&0.28&\\
model 1a&0.99&0.08&1.91&0.55&49.62\\
model 1b&1&0.09&2.17&0.49&42.81\\
model 1c&0.98&0.09&1.97&0.52&47.71\\
model 1d&1&0.11&3.05&0.35&19.67\\
model 2a&0.99&0.08&1.92&0.55&49.47\\
model 2b&1&0.09&2.17&0.49&42.88\\
model 2c&0.97&0.09&1.96&0.53&48.07\\
model 2d&1&0.1&3.04&0.35&19.74\\
\hline
\end{tabular}
\label{results-sc1}
\caption{Comparison of models based on simulation scenario 2 by treatment contrast.}
\end{table}

\end{document}